\documentclass[]{interact}

\usepackage{epstopdf}
\usepackage[caption=false]{subfig}

\usepackage[numbers,sort&compress]{natbib}
\usepackage{hyperref}

\bibpunct[, ]{[}{]}{,}{n}{,}{,}
\makeatletter
\def\NAT@def@citea{\def\@citea{\NAT@separator}}
\makeatother

\theoremstyle{plain}

\theoremstyle{definition}

\theoremstyle{remark}

\begin{document}


\title{Topological electronic bands in crystalline solids}

\author{
\name{A.~T. Boothroyd$^{\textrm{a},\textrm{b}}$\thanks{CONTACT Andrew T.~Boothroyd. Email: andrew.boothroyd@physics.ox.ac.uk}}
\affil{$^\textrm{a}$Department of Physics, Oxford University, Clarendon Laboratory, Oxford, UK}
\affil{$^\textrm{b}$LINXS Institute of Advanced Neutron and X-ray Science, Lund, Sweden}
}

\maketitle

\begin{abstract}
Topology is now securely established as a means to explore and classify electronic states in crystalline solids.  This review provides a gentle but firm introduction to topological electronic band structure suitable for new researchers in the field. I begin by outlining the relevant concepts from topology, then give a summary of the theory of non-interacting electrons in periodic potentials. Next, I explain the concepts of the Berry phase and Berry curvature, and derive key formulae. The remainder of the article deals with how these ideas are applied to classify crystalline solids according to the topology of the electronic states, and the implications for observable properties. Among the topics covered are the role of symmetry in determining band degeneracies in momentum space, the Chern number and $\mathcal{Z}_2$ topological invariants, surface electronic states, two- and three-dimensional topological insulators, and Weyl and Dirac semimetals
\end{abstract}


\section{Introduction}

Over the past few decades, topology has become an increasingly valuable tool in condensed matter physics, providing a radically different way of thinking about materials and enabling the prediction and discovery of exotic states with intriguing physical properties. The importance of topology in condensed matter was recognised by the award of the 2016 Nobel Prize in Physics to Thouless, Haldane and Kosterlitz.

Amongst topological themes of current interest in the area of quantum materials are, in no particular order, topological defects and phase transitions \cite{Kosterlitz2017, Haldane2017}, topological insulators \cite{KaneMoore2011}, graphene \cite{Pachos2009}, magnetic skyrmions \cite{Lancaster2019}, and Majorana fermions for quantum computation \cite{AguadoKouwenhoven2020}.  The use of topology is enriching other fields too, including photonics \cite{Segev-Bandres2021}, polymer statistics \cite{GuZhaoJohnson2019}, and biological matter \cite{Ardaseva2022}. Topological systems are also appealing from the point of view of practical applications, due in large part to a special property known as \emph{topological protection}, which can enhance the stability of a state.

In short, concepts from topology have become widespread in condensed matter and are making a significant impact. My focus here is on one particular area within this new field, namely the topology of electronic bands in crystalline solids and associated observable phenomena. The topics I shall review came to prominence in the early years of the twenty-first century, first with the demonstration of the extraordinary electronic properties of graphene \cite{Novoselov2005}, and very shortly after with the prediction \cite{KaneMele2005a,Bernevig2006} and subsequent experimental verification \cite{Konig2007} of the quantum spin Hall insulator (a two-dimensional version of a topological insulator). These ideas were soon extended into three dimensions, igniting the field of topological electronic bands in solids and leading to a range of predictions and discoveries of novel effects in topological insulators, semimetals, metals and superconductors.

Several important themes cut across this new realm of electronic phenomena:

\begin{enumerate}
\item The \emph{Berry phase}. This is a quantum-mechanical phase picked up by systems when they are slowly deformed. Berry phases cause interference effects due to the wave-like nature of particles.

\item The \emph{Berry curvature}. This is an emergent gauge field which acts like a magnetic field in momentum space and gives rise to unconventional transport phenomena, such as the anomalous Hall and Nernst effects.

\item \emph{Band touching points}. These are points in momentum space where two or more electronic bands become degenerate. They are special because of the associated Berry phase effects, and also because if the chemical potential is located near such a point the electronic excitations can be exotic.
\item \emph{Electronic topology}. The materials we are concerned with here contain topologically non-trivial electronic states. In other words, the geometry of electrons in the solid is distinct from the geometry of a free electron. Topologically inequivalent states cannot be smoothly deformed into one another, and are often separated by an energy barrier, making them stable against perturbations.
\end{enumerate}

Accounts of the above-mentioned concepts are plentiful but can be somewhat technical. My aim here is to provide a mathematically rigorous but pedagogical introduction, and to show as simply as possible how these ideas translate into observable phenomena. Some proficiency is quantum mechanics and solid-state physics is required, but not beyond what is taught in a standard undergraduate course. For readers who aspire to know more, there are some excellent books and technical review articles which go into much greater detail, of which Refs.~\cite{MoessnerMoore,Vanderbilt,Bernevig,Kane2013,CayssolFuchs2021,Armitage2018,XiaoChiangNiu2010,Yang2016,Lv2021,Rachel2018,Tokura2019,WehlingSchafferBalatsky2014,GeimNovoselov2007,SatoAndo2017} are a good selection.

\section{Basic ideas of topology}\label{sec:topology}

I begin with a brief introduction to a few important concepts in topology that are relevant to the theme of this article.

\subsection{Topological invariants}\label{sec:topological-invariant}

\begin{figure}
\vspace*{-1cm}
\setlength{\abovecaptionskip}{0pt plus 0pt minus 0pt}
\centering
\includegraphics[width=0.5\textwidth, angle=90]{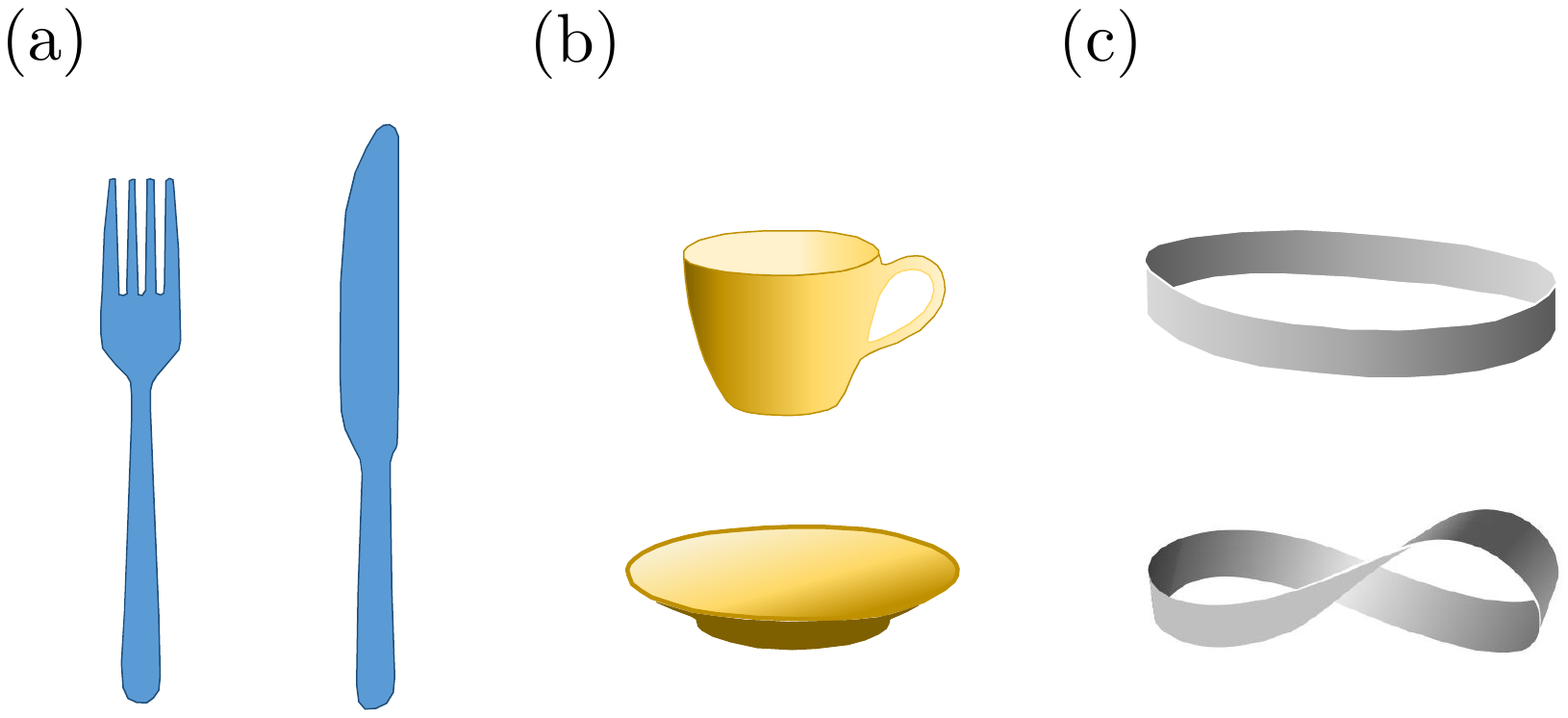}
\caption{Examples of pairs of objects which are (a) topologically equivalent, and (b)--(c) not topologically equivalent.} \label{fig:homeomorphism}
\end{figure}

Topology is first and foremost a branch of mathematics used to describe geometrical properties that remain invariant when an object is continuously deformed. Two objects are considered to be topologically equivalent, or \emph{homeomorphic}, if one object can be transformed into the other by bending, stretching or twisting, but not cutting or sticking together. For example, a knife and fork are homeomorphic, Figure~\ref{fig:homeomorphism}(a), whereas a cup and saucer are not, Figure~\ref{fig:homeomorphism}(b).  Similarly, to convert a simple cylindrical band into a M\"{o}bius strip, Figure~\ref{fig:homeomorphism}(c), you need to cut the band, apply a half-twist, and then reattach the ends, so these shapes are not topologically equivalent.

In topology, objects are classified by means of \emph{topological invariants}, which quantify some generic property which they have in common, e.g.~the number of holes or twists. Topological invariants have discrete values, usually normalised to be integers. A well-known example is Euler's characteristic for geometric solids, defined as $\chi = F + V - E$, where $F$, $V$ and $E$ are the number of faces, vertices and edges. All convex polyhedra have $\chi = 2$, and are topologically equivalent to (and hence continuously deformable into) a sphere. Other examples of topological invariants are, winding numbers (of closed curves) and the Kauffman bracket polynomial (for knots).

Topological equivalence in quantum systems concerns whether the Hamiltonian of one system can be continuously deformed into that of another while preserving a topological invariant (also known as a topological quantum number). Here, `deforming' means slowly changing one or more of the parameters of the Hamiltonian. The topology of electronic band states in crystals is characterised by a topological invariant called the \emph{Chern number}, which I shall define in Section~\ref{sec:Chern-theorem}.

\subsection{Topological protection}\label{sec:topological-protection}

 We know from experience that it is often difficult to convert topologically inequivalent states into one another.  A knot, for example, cannot easily be removed from a piece of string simply by shaking it around. It requires a series of coordinated steps that are highly unlikely to occur at random. It is easy to appreciate, therefore, that a state which is topologically distinct from its surroundings can be highly immune to perturbations, making it especially robust. States which are topologically protected in this way can be very useful in devices. For example, topologically protected spin-polarised surface conducting states have potential applications in high-speed electronics and spintronics, and an entirely new field of topological quantum computation based on the manipulation of Majorana edge modes is envisaged.

\subsection{Topological order}\label{sec:topological-order}

 Topologically inequivalent states are regarded as being differently ordered, but not in a conventional sense. Topological electronic order is characterised by a Chern number, rather than by a broken symmetry such as loss of rotation, inversion or time-reversal symmetry. When topological transitions take place between states with different Chern numbers something dramatic must happen --- something like a twisting of the electronic wave function. One consequence is the existence of edge modes at the boundary between different topological phases. An example is the conducting states that exist on the surface of a topological insulator.

\subsection{What is a topological insulator?}\label{sec:topological-insulators}

\begin{figure}
\vspace*{-1.5cm}
\hspace*{-2cm}
\setlength{\abovecaptionskip}{-150pt plus 0pt minus 0pt}
\centering
\includegraphics[width=0.9\textwidth, angle=90]{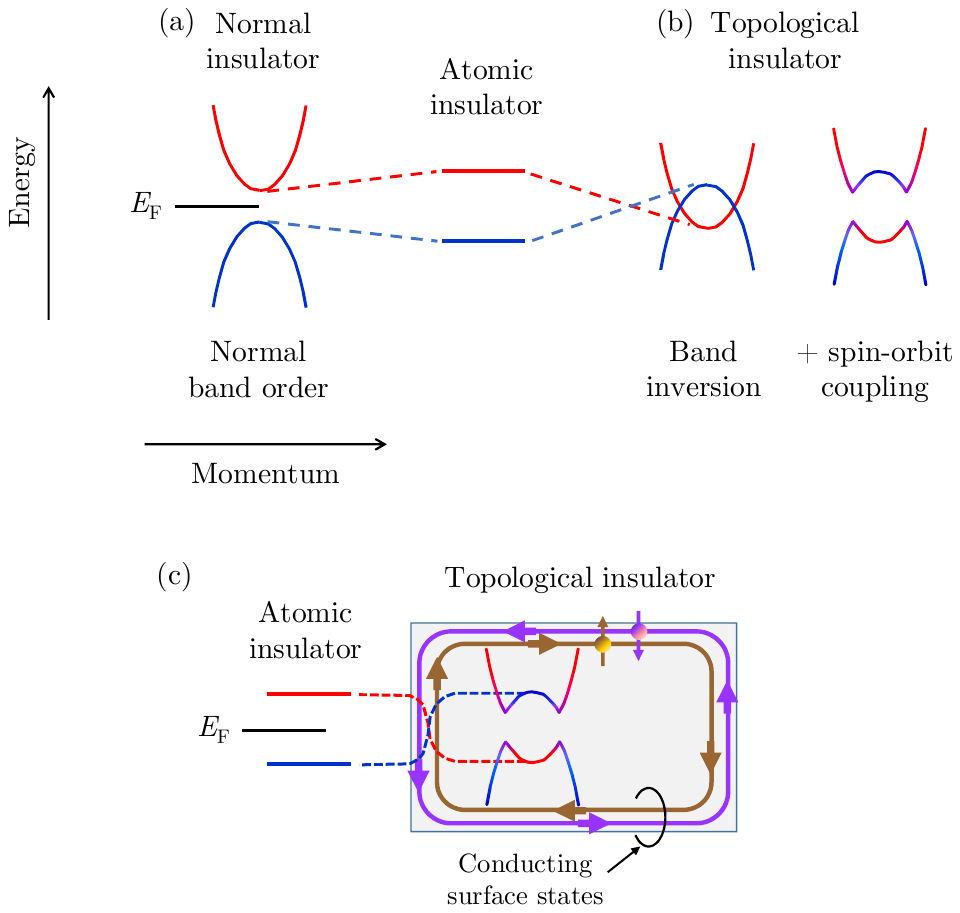}
\setlength{\abovecaptionskip}{2pt plus 0pt minus 0pt}
\caption{Illustration of band inversion and edge currents in a topological insulator. For simplicity, two atomic levels are shown, the lower level being occupied and the upper level empty, so the Fermi level ($E_\textrm{F}$) is in the gap. (a) The band gap of a normal insulator is smoothly connected to the gap between the atomic levels. (b) In a topological insulator there is a gap between the valence and conduction bands, and for some range of momenta the bands are inverted relative to the atomic levels. (c) Gap closure and level crossing occurs at the interface between the topological insulator and vacuum, resulting in a conducting surface. The surface currents are spin-polarised. The up-spin and down-spin currents flow in opposite directions around the boundary.} \label{fig:TI}
\end{figure}

The classification of electronic states in crystalline solids according to their topology was pioneered by Thouless and coworkers, who in 1982 identified the topological invariant for the integer quantum Hall effect \cite{TKNN1982}. It would take more than twenty years for this way of thinking to become mainstream. Theoretical investigations on quantum spin Hall insulators \cite{KaneMele2005a,KaneMele2005b,Bernevig2006}, now more commonly known as two-dimensional topological insulators, led to the first experimental observations, reported in 2007, on HgTe/CdTe quantum wells \cite{Konig2007}. The concepts were soon extended to describe three-dimensional topological insulators \cite{FuKaneMele2007,MooreBalents2007,Roy2009}, which were identified in experiments shortly afterwards \cite{Hsieh2008,Xia2009}.

The concept of the topological insulator (TI) is actually very simple. Figure~\ref{fig:TI} illustrates the basic idea. The energy levels of isolated atoms are quantised. When the atoms come together to form a solid the levels spread out to form a band of states whose energy varies (disperses) with momentum. An insulator is a material with an energy gap that separates the (filled) valence bands from the (empty) conduction bands. If, as the atoms come together and bands form, the gap between the valence and conduction bands never closes, then we have a trivial (i.e.~non-topological) insulator or semiconductor. This situation is illustrated in Figure~\ref{fig:TI}(a) for a single pair of levels.

Alternatively, if the levels cross as the solid forms, then for at least some states the energies will be inverted compared with the isolated atom. If, in addition, a gap forms in the solid where the lower and upper atomic levels cross, then we still have an insulator in the bulk, but now it is a topological insulator, Figure~\ref{fig:TI}(b).  The reason why this insulator is topological is that when the levels cross, the valence and conduction bands twist, metaphorically speaking, a bit like going from the simple cylindrical band to the M\"{o}bius strip illustrated in Figure~\ref{fig:homeomorphism}(c). Put another way, the insulating states in the solid are not smoothly connected to the atomic states.

One interaction which can open a gap where the levels cross in the bulk is spin--orbit coupling (SOC). Accordingly, topological insulators are often made from elements with high atomic numbers (hence large SOC). Despite the bulk gap, the surface (or edge in two dimensions) of a topological insulator must necessarily be conducting in the absence of a magnetic interaction. In simple terms, the states which are inverted in the bulk must cross back to their usual band ordering on moving from inside to outside the solid. When they cross, the energy gap closes and so charge can flow. This band crossing happens just beneath the surface, where the bulk lattice periodicity of the material is terminated, Figure~\ref{fig:TI}(c). Gapless surface or edge states are then inevitable. Their existence depends only on there being a transition between two topologically inequivalent states below and above the surface. This aspect of topological materials is referred to as the \emph{bulk--boundary correspondence}.

More detailed analysis shows that the boundary states come in Kramers degenerate pairs (see Section~\ref{subsec:Kramers}) which are spin-polarised, meaning that states with opposite spin move in opposite directions along a given boundary, see Figure~\ref{fig:TI}(c).  In consequence, these states cannot be backscattered by non-magnetic impurities, because no conduction channel exists for electrons whose direction of motion is reversed but whose spin is unchanged (scattering by non-magnetic impurities preserves the electron spin direction). The flow of spin current, therefore, is dissipationless, which is highly advantageous for low-power spintronics applications. Further exotic properties emerge when TIs are coupled to electric and magnetic fields, or to superconductors \cite{Qi2009,QiHughesZhang2008,FuKane2008}.

In summary, topological insulators are insulating in the bulk but metallic on the surface, carry dissipationless surface spin currents, and are topologically distinct from a normal insulator.

\section{Electronic band structure in crystalline solids}\label{sec:band-structure}

In this section I review some of the features of electronic states in crystalline solids, as first considered almost a century ago by Felix Bloch in his PhD thesis \cite{BlochZPh29}. We are going to be concerned primarily with systems in which the electrons can be described adequately as non-interacting particles, and for the time being I shall neglect spin.

\subsection{Lattice Hamiltonian}\label{subsec:lattice}

The single-particle states are solutions of the time-independent Schr\"{o}dinger equation
\begin{equation}
{\mathcal H}\psi(\textbf{r}) = E\psi(\textbf{r}).
\label{eq:TISE-Bloch}
\end{equation}
The Hamiltonian $\mathcal H$  contains the crystal potential $V(\textbf{r})$ which satisfies the periodicity condition
\begin{equation}
V(\textbf{r}+\textbf{R}) = V(\textbf{r})
\label{eq:periodic-potential}
\end{equation}
for any lattice vector
\begin{equation}
\textbf{R} = n_1\textbf{a}_1 + n_2\textbf{a}_2 + n_3\textbf{a}_3\ \ \ \ \ \ (n_1, n_2, n_3\ \textrm{integers}),
\label{eq:lattice-vector}
\end{equation}
where $\textbf{a}_1$, $\textbf{a}_2$ and $\textbf{a}_3$ are the primitive lattice vectors.

\begin{figure}
\vspace*{-3.0cm}
\setlength{\abovecaptionskip}{-30pt plus 0pt minus 0pt}
\centering
\includegraphics[width=0.6\textwidth, angle=90]{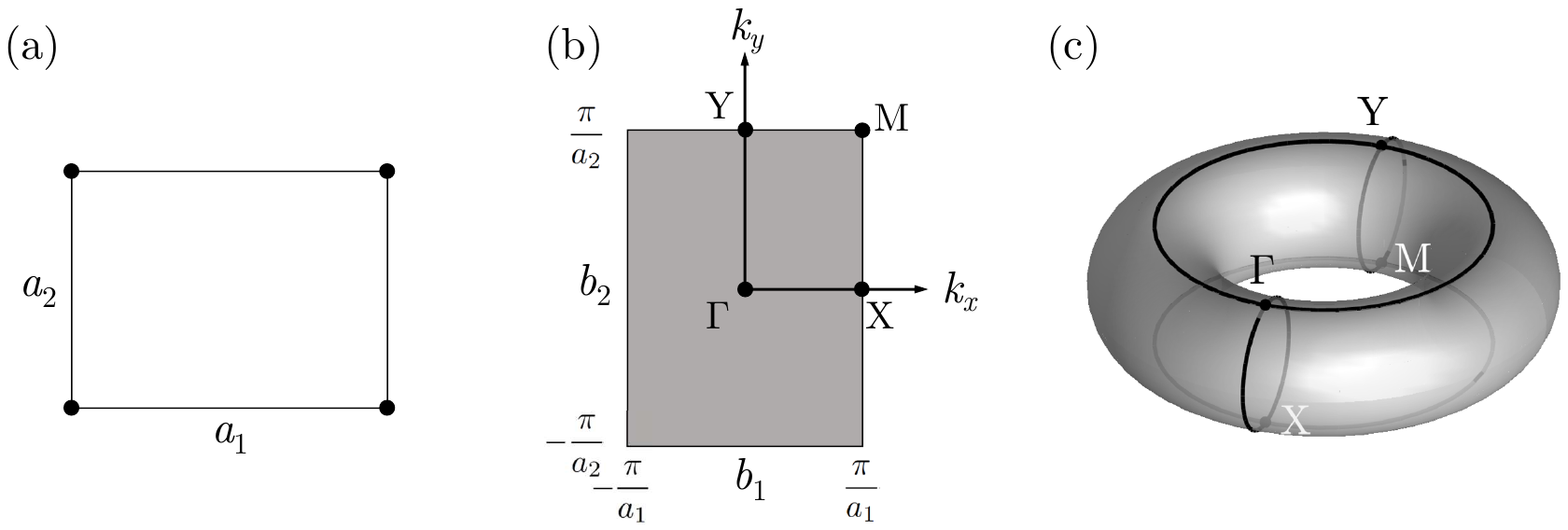}
\caption{(a) Rectangular lattice with primitive lattice vectors $a_1$ and $a_2$. (b) Wigner--Seitz cell (Brillouin zone) centred on $k_x = k_y = 0$  for the rectangular lattice in (a), with primitive reciprocal lattice vectors $b_1 = 2\pi/a_1$ and $b_2 = 2\pi/a_2$. (c) Toroidal geometry of the BZ when the Bloch functions are defined on the periodic gauge. The rectangular BZ in (b) is first rolled around the $k_y$ axis, and then the edges at $k_y = \pm\frac{\pi}{a_2}$ are joined together to form a torus.} \label{fig:BZ}
\end{figure}

The Fourier transform of the real-space lattice defines the \emph{reciprocal lattice}, which is an essential tool for describing the properties of waves propagating in a periodic potential. The reciprocal lattice is the set of discrete vectors
\begin{equation}
\textbf{G} = m_1\textbf{b}_1 + m_2\textbf{b}_2 + m_3\textbf{b}_3\ \ \ \ \ \ (m_1, m_2, m_3\ \textrm{integers}),
\label{eq:RLV}
\end{equation}
where $\textbf{b}_i = (2\pi/v_0)\, \textbf{a}_j \times \textbf{a}_k$ are primitive reciprocal lattice vectors with the property $\textbf{a}_i\cdot \textbf{b}_j = 2\pi\delta_{ij}$. Here, $v_0$ is the volume of the primitive unit cell of the direct lattice and $\delta_{ij}$ is the Kronecker delta. Panels (a) and (b) of Figure~\ref{fig:BZ} illustrate the unit cell of a primitive rectangular lattice and the corresponding reciprocal lattice.

\subsection{Bloch functions}\label{subsec:Bloch-fns}

\begin{figure}
\centering
\includegraphics[width=0.75\textwidth, angle=0]{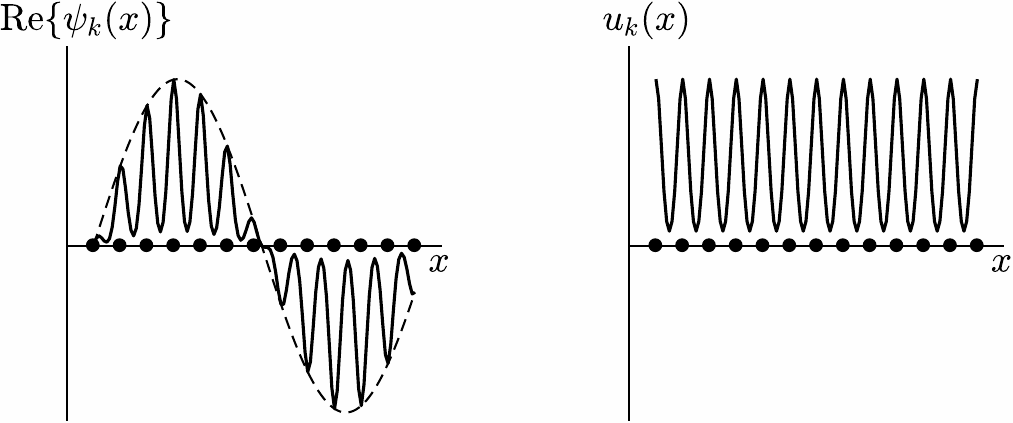}\vspace{1.2cm}
\caption{Representation of a one-dimensional Bloch function. The real part of the Bloch function is plotted on the left, and the lattice-periodic function is shown on the right. A value $k = 0.5/a$ has been used, where $a$ is the spacing between the lattice points (represented by dots).} \label{fig:Bloch}
\end{figure}

Bloch showed that the solutions of Equations~(\ref{eq:TISE-Bloch})--(\ref{eq:lattice-vector}) have the generic form (see Figure~\ref{fig:Bloch})
\begin{align}
\psi_\textbf{k}(\textbf{r}) = \textrm{e}^{i\textbf{k}\cdot\textbf{r}}u_\textbf{k}(\textbf{r}),
\label{eq:Bloch-fn}
\end{align}
which comprises a plane wave modulated by a function $u_\textbf{k}(\textbf{r})$ that is periodic in the lattice,
\begin{equation}
u_\textbf{k}(\textbf{r}+\textbf{R}) = u_\textbf{k}(\textbf{r}).
\label{eq:periodicity-BF}
\end{equation}
The \emph{Bloch wavevector} \textbf{k} resembles the wavevector of a free particle, but differs in two important respects. First, it is quantised due to the confinement of the particle in the solid. Second, translational symmetry makes \textbf{k} states that differ by a reciprocal lattice vector equivalent, in the sense that $\psi_\textbf{k}(\textbf{r})$ and $\psi_{\textbf{k}+\textbf{G}}(\textbf{r})$ correspond to the same energy eigenstate for any reciprocal lattice vector \textbf{G}.

These two properties mean that the \textbf{k} vectors which label unique solutions of the Schr\"{o}dinger equation are finite in number and can be contained in a bounded volume of \textbf{k} space. This volume is known as the Brillouin zone (BZ), and is usually chosen to be the Wigner--Seitz unit cell\footnote{The Wigner--Seitz unit cell is defined as the region of \textbf{k} space centred on a reciprocal lattice point $\Gamma$ such that it contains all \textbf{k} states which are closer to $\Gamma$ than to any other reciprocal lattice point. } centred on $\textbf{k} = 0$. The BZ of a primitive rectangular lattice is illustrated in Figure~\ref{fig:BZ}.

The fact that \textbf{k} and $\textbf{k}+\textbf{G}$ correspond to the same energy eigenstate means that the Bloch functions $\psi_\textbf{k}(\textbf{r})$ and $\psi_{\textbf{k}+\textbf{G}}(\textbf{r})$ can differ at most by a trivial phase factor $\textrm{e}^{-i\theta}$. In what follows we shall choose $\theta = 0$, a condition known as the \emph{periodic gauge}. With this choice of phase the BZ has periodic boundary conditions on all its faces, giving it the topology of a $d$-torus in $d$ dimensions, i.e.~a circle in one dimension, the surface of a doughnut in two dimensions, and a 3-torus in three dimensions. Figure~\ref{fig:BZ}(c) illustrates the 2-torus geometry of the BZ in two dimensions.

The eigenvalues $E_\textbf{k}$ associated with the Bloch functions form a band of states in \textbf{k} space which originates from the overlap of atomic orbitals.   The degree of orbital overlap in the solid determines the electronic band width. Atoms have multiple orbitals, so for a given \textbf{k} there will be multiple energy levels $E_{n\textbf{k}}$.  The additional quantum number $n$, called the band index, is used to label the different bands. Each energy band has its own characteristic variation with \textbf{k}, and different bands can cross over one another. Band crossing points play an important r\^{o}le in the phenomena we shall discuss here.
.

\section{The Berry phase}\label{sec:Berry-phase}

\subsection{Quantum mechanical phase factors}\label{subsec:QM-phases}

Phase factors have already been mentioned, and are an important feature of quantum mechanical functions. They take the form $\textrm{e}^{-i\theta}$, where $\theta$ is a (real-valued) phase angle. Some phase factors are trivial.  For example, suppose that $\psi$ is a solution of the time-independent Schr\"{o}dinger equation, Equation (\ref{eq:TISE-Bloch}). The gauge transformation $\psi' = \textrm{e}^{-i\beta}\psi$, where $\beta$ commutes with $\mathcal H$, simply generates an equivalent solution $\psi'$ with the same energy eigenvalue. There are no observable consequences of such a gauge transformation.

Another example is the dynamical phase of the eigenfunction, which comes from the time-dependent Schr\"{o}dinger equation,
\begin{equation}
{\mathcal H}\psi = i\hbar\frac{\partial\psi}{\partial t}.
\label{eq:TDSE}
\end{equation}
Substituting (\ref{eq:TISE-Bloch}) in (\ref{eq:TDSE}) and integrating for a time-independent ${\mathcal H}$, we obtain
\begin{equation}
\psi(t) = \textrm{e}^{-iEt/\hbar} \psi(0),
\label{eq:dynamical-phase-1}
\end{equation}
or, more generally,
\begin{equation}
\psi(t) = \textrm{e}^{-i\int E(t) \textrm{d}t/\hbar} \psi(0).
\label{eq:dynamical-phase-2}
\end{equation}
We see that the time dependence of the eigenfunction is contained in a time-dependent phase factor.

\subsection{Adiabatic changes and the Berry phase}\label{subsec:adiabatic}

The Berry phase is a \emph{non-trivial} phase that can have observable consequences, and is the central idea of topological band theory. The Berry phase arises when a quantum state is varied adiabatically, through slow changes in the parameters on which the state depends. As we shall show, variations in the Berry phase are particularly strong in the vicinity of points where electronic levels become degenerate.

\begin{figure}
\vspace*{-2cm}
\setlength{\abovecaptionskip}{-10pt plus 0pt minus 0pt}
\centering
\includegraphics[width=0.6\textwidth, angle=90]{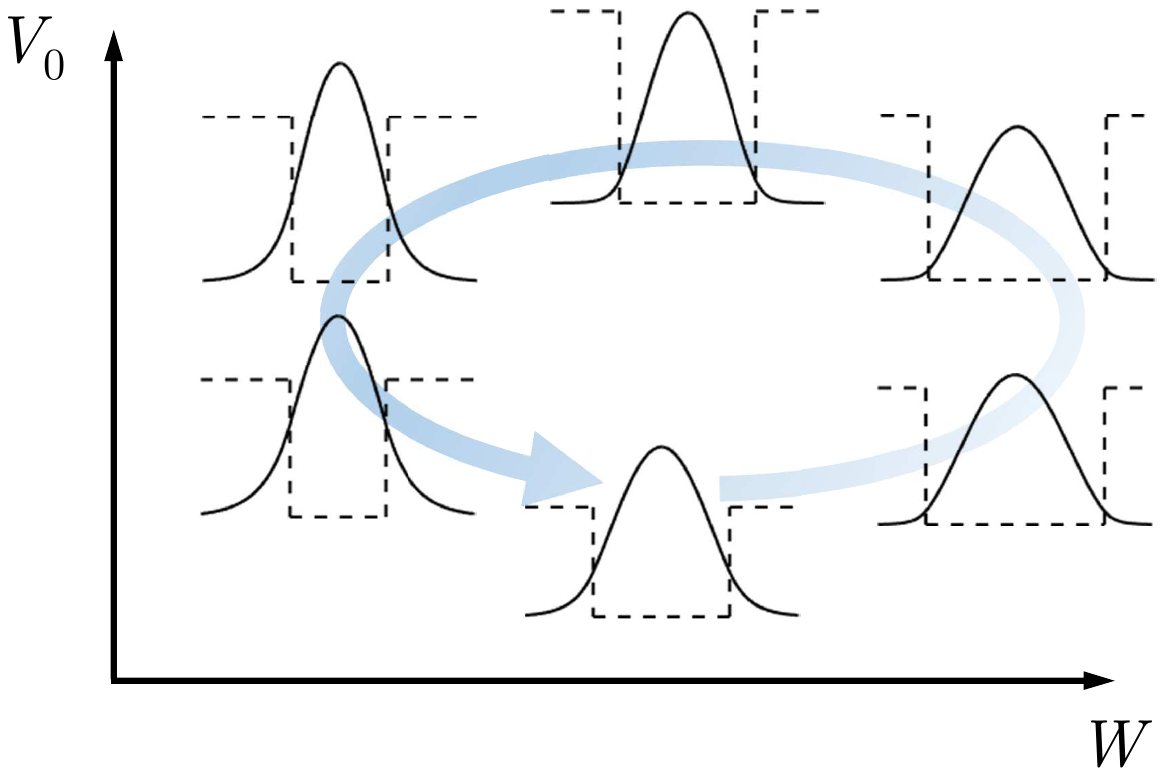}
\caption{Illustration of the adiabatic theorem in quantum mechanics. The plot shows the probability density $|\psi^2|$ for the ground state wave function of a particle in a one-dimensional potential well, as the width $W$ and depth $V_0$ of the well are varied around a closed path. If the parameters are changed sufficiently slowly then the wave function adapts so that the particle remains in the instantaneous ground state throughout the cycle.} \label{fig:1Dwell}
\end{figure}

In his original work \cite{BerryPRS84}, Berry considered a general system whose Hamiltonian depends on a set of parameters that can be varied as a function of time. He showed that if the system starts in an eigenstate $\psi(0)$ of the Hamiltonian, and the system is taken slowly around a closed path in parameter space, then at the end of the cycle the system returns to $\psi(0)$ but acquires a phase factor $\textrm{e}^{-i\gamma}$ \emph{in addition to} the usual dynamical phase factor given in Equation~(\ref{eq:dynamical-phase-2}).  The phase angle $\gamma$, now known as the \emph{Berry phase}, depends upon how the eigenstates of the Hamiltonian change around the path. In other words, the Berry phase is a function of the geometry of the Hamiltonian in parameter space.

The type of process we are considering is described by the \emph{adiabatic theorem} in quantum mechanics, and is illustrated in Figure~\ref{fig:1Dwell}. The picture shows how the ground-state of a particle in a one-dimensional square potential well evolves as the width and depth of the well are varied around a closed path. If the changes occur sufficiently slowly, then the wave function has time to adapt and the particle remains in the instantaneous ground state throughout the process. By `sufficiently slowly', we mean that the characteristic time-scale for the process (e.g.~the time to complete the cycle) must be large compared with $\hbar/\Delta E$, where $\Delta E$ is the separation between the ground state and the rest of the spectrum of eigenstates.

\subsection{Cell-periodic Bloch functions and the Bloch Hamiltonian}\label{subsec:Bloch-H}

We are interested in results that apply to systems of non-interacting electronic quasiparticles in periodic potentials.  The parameter space in which adiabatic changes take place is in this case the space spanned by \textbf{k}, which for present purposes can be regarded as a quasi-continuous variable rather than a discrete quantum number.\footnote{The number of allowed \textbf{k} states in the first BZ is equal to the number of primitive unit cells in the crystal, so for a macroscopic system the allowed \textbf{k} are very close together relative to the size of the BZ.} To emphasise the role of \textbf{k} as a (quasi-continuous) parameter we shall from now on write the \textbf{k}-dependence as an argument rather than as a subscript, e.g.~$E_n(\textbf{k})$ instead of $E_{n\textbf{k}}$.

In what follows it will be necessary to work with the lattice-periodic part of the Bloch function (see Section~\ref{sec:band-structure}), which is also known as the \emph{cell-periodic} Bloch function,
\begin{equation}
u_n(\textbf{k}) = \textrm{e}^{-i\textbf{k}\cdot\textbf{r}}\psi_n(\textbf{k}).
\label{eq:cell-per-BF}
\end{equation}
The energy eigenvalue equation (\ref{eq:TISE-Bloch}) may then be written
\begin{align}
{\mathcal H}(\textbf{k})u_n(\textbf{k}) = E_n(\textbf{k})u_n(\textbf{k}),
\label{eq:TISE-Hk}
\end{align}
where we have defined the \emph{Bloch Hamiltonian}
\begin{equation}
{\mathcal H}(\textbf{k}) = \textrm{e}^{-i\textbf{k}\cdot\textbf{r}}\,{\mathcal H}\,\textrm{e}^{i\textbf{k}\cdot\textbf{r}}.
\label{eq:Bloch-H}
\end{equation}
The cell-periodic Bloch functions are seen to be eigenstates of ${\mathcal H}(\textbf{k})$, and form a complete set of basis states defined on the primitive unit cell. Hereafter, we shall adopt Dirac's notation and represent the basis state $u_n(\textbf{k})$ by the ket $|u_n(\textbf{k})\rangle$.

\subsection{Derivation of the Berry phase and Berry connection}\label{subsec:BerryPhaseConn}

Having dispensed with the preliminaries, we now come to the essence of the problem which is to obtain expressions for the Berry phase and related quantities when electronic quasiparticles move slowly around a closed path $\mathcal C$ in \textbf{k} space. A particular physical significance of the calculations is that they apply to the motion of quasiparticles in electric and/or magnetic fields under many conditions of interest, and therefore describe a variety of magnetotransport phenomena.

To proceed, we write the state of the system at time $t$, when it has reached a point \textbf{k} on the path, as
\begin{equation}
|\psi(t)\rangle = \textrm{e}^{-i\theta(t)}|u_n(\textbf{k})\rangle,
\label{eq:Berry-phase-0}
\end{equation}
where $|u_n(\textbf{k})\rangle$ is one of the normalised eigenstates of ${\mathcal H}(\textbf{k})$, see Equation~(\ref{eq:TISE-Hk}). Consistent with the adiabatic theorem, we assume that the system remains in the same quantum level $n$ (in other words, in the same band) throughout the process.  Substituting $|\psi(t)\rangle$ into the Schr\"{o}dinger Equations~(\ref{eq:TISE-Bloch}) and (\ref{eq:TDSE}), and taking the inner product\footnote{The inner product is an integral over the unit cell in real space.}  with $\langle u_n(\textbf{k})|$, we obtain
\begin{equation}
\hbar\frac{\textrm{d}\theta}{\textrm{d}t} = E_n(t) - i\hbar \langle u_n(\textbf{k})|\frac{\textrm{d}}{\textrm{d}t}|u_n(\textbf{k})\rangle.
\label{eq:Berry-phase-2}
\end{equation}
Hence, if it takes time $T$ to complete the full closed path, then
\begin{equation}
\theta(T) = \frac{1}{\hbar}\int_0^T E_n(t) \textrm{d}t - i \int_0^T\langle u_n(\textbf{k})|\frac{\textrm{d}}{\textrm{d}t}|u_n(\textbf{k})\rangle\,\textrm{d}t.
\label{eq:Berry-phase-3}
\end{equation}
The first term is the usual dynamical phase --- see Equation~(\ref{eq:dynamical-phase-2}). The second term defines the \emph{Berry phase},
\begin{equation}
\gamma_n =  i \int_0^T\langle u_n(\textbf{k})|\frac{\textrm{d}}{\textrm{d}t}|u_n(\textbf{k})\rangle\,\textrm{d}t.
\label{eq:Berry-phase-4}
\end{equation}
 The Berry phase is always contained in a phase factor $\textrm{e}^{-i\gamma_n}$ in expressions that relate to observable quantities, and so values of $\gamma_n$ that differ by an integer multiple of $2\pi$ are the same for all practical purposes. In other words, the Berry phase is defined modulo $2\pi$. This property has important implications for band topology, as will be discussed in Section~\ref{sec:Chern-theorem}.

Time can be eliminated from (\ref{eq:Berry-phase-4}) via the chain rule,
\begin{equation}
\frac{\textrm{d}}{\textrm{d}t} = \nabla_\textbf{k}\cdot\frac{\textrm{d}\textbf{k}}{\textrm{d}t},
\label{eq:chain-rule}
\end{equation}
where $\nabla_\textbf{k}$ stands for $(\frac{\partial}{\partial k_x}, \frac{\partial}{\partial k_y}, \frac{\partial}{\partial k_z})$, whereupon
\begin{equation}
\gamma_n =  i \oint_\mathcal{C} \langle u_n(\textbf{k})|\nabla_\textbf{k}|u_n(\textbf{k})\rangle\cdot\textrm{d}\textbf{k}.
\label{eq:Berry-phase-5}
\end{equation}
The matrix element $\langle u_n|\nabla|u_n\rangle$ is an imaginary number, so the Berry phase is always real.\footnote{This can be seen by applying $\nabla_\textbf{k}$ to the normalization condition $\langle u_n|u_n\rangle = 1$: \begin{align}\nabla\langle u_n|u_n\rangle = \langle \nabla u_n|u_n\rangle + \langle u_n|\nabla u_n\rangle = \langle u_n|\nabla u_n\rangle^\ast + \langle u_n|\nabla u_n\rangle = 0,\nonumber\end{align} hence $\langle u_n|\nabla u_n\rangle = -\langle u_n|\nabla u_n\rangle^\ast$ is purely imaginary and therefore $\gamma_n$ is purely real. } The integrand in Equation~(\ref{eq:Berry-phase-5}) defines the \emph{Berry connection} (or \emph{Berry potential}) $\boldsymbol{\mathcal{A}}_n(\textbf{k})$,
\begin{equation}
\boldsymbol{\mathcal{A}}_n(\textbf{k}) =  i \langle u_n(\textbf{k})|\nabla_\textbf{k}| u_n(\textbf{k})\rangle,
\label{eq:Berry-connection}
\end{equation}
so that
\begin{equation}
\gamma_n =   \oint_\mathcal{C} \boldsymbol{\mathcal{A}}_n(\textbf{k})\cdot \textrm{d}\textbf{k}.
\label{eq:Berry-phase-6}
\end{equation}
As we shall see in Equation~(\ref{eq:Berry-curvature-curl}), $\boldsymbol{\mathcal{A}}_n(\textbf{k})$ is analogous to the magnetic vector potential in electromagnetism. One property they have in common is that both potentials are gauge dependent. To see this, we apply a gauge transformation to $|u_n(\textbf{k})\rangle$:
\begin{align}
|u_n'(\textbf{k})\rangle & =\textrm{e}^{-i\beta(\textbf{k})}|u_n(\textbf{k})\rangle\nonumber\\
\rightarrow\ \  \boldsymbol{\mathcal{A}}_n' & = \boldsymbol{\mathcal{A}}_n + \nabla_\textbf{k} \beta(\textbf{k}).
\label{eq:Berry-connection-gauge-dep}
\end{align}
An observable quantity cannot depend on an arbitrary choice of phase $\beta(\textbf{k})$, so $\boldsymbol{\mathcal{A}}_n$ is not an observable quantity.

\subsection{Parallel transport}\label{subsec:Parallel}

\begin{figure}
\vspace*{-3.5cm}
\setlength{\abovecaptionskip}{-50pt plus 0pt minus 0pt}
\centering
\includegraphics[width=0.6\textwidth, angle=0]{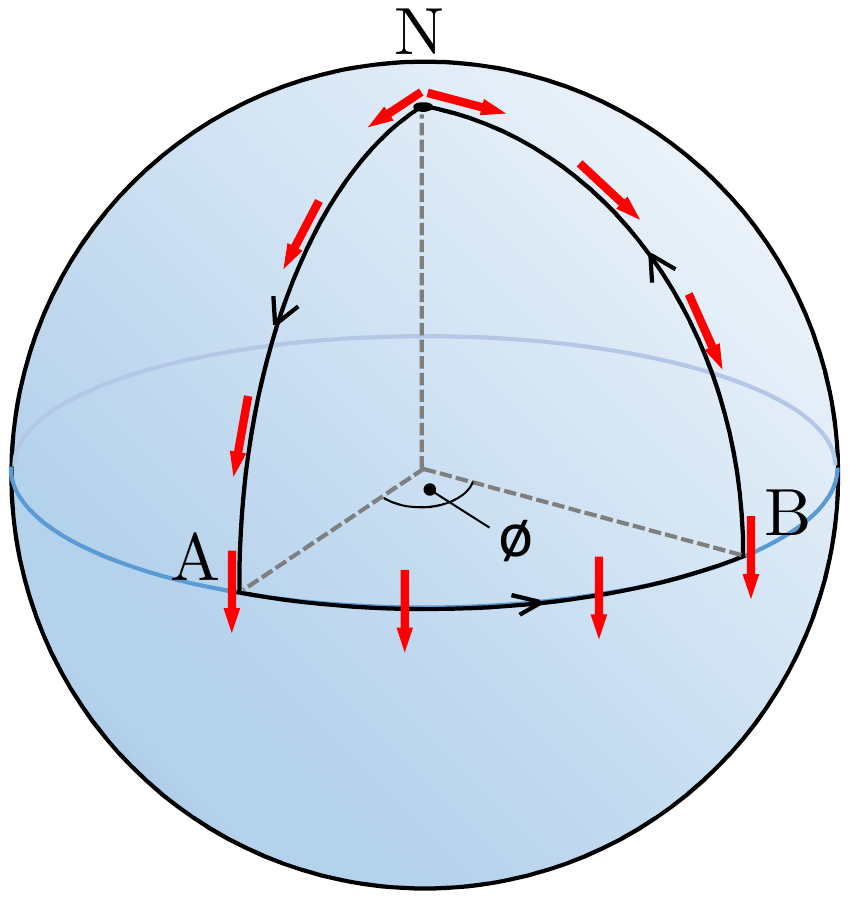}\vspace{0cm}
\caption{Parallel transport of a vector on a sphere. The vector travels around the closed path N--A--B--N while remaining parallel to the surface of the sphere. The angle between the initial and final directions of the vector is $\phi$. } \label{fig:Parallel}
\end{figure}

Let us pause for a moment. The Berry phase is an example of a phenomenon called \emph{anholonomy}, in which a system variable fails to return to its original value when the system is driven by other variables around a cycle.  A simple classical example is the parallel transport of a vector (e.g.~an arrow) on a sphere, Figure~\ref{fig:Parallel}. Starting at the north pole (N), an arrow is transported down a meridian until it reaches the equator (A), at which point it follows the equator to point B, then returns back to N along the meridian through B. The arrow remains parallel to the local surface at all points on the path by the minimum of adjustments to its orientation. At the end of the journey, the arrow is seen to point in a different direction than at the start of the journey.

In this example, the parameters of the path are the coordinates on the surface of the sphere, and the anholonomy occurs because of the curved geometry of the parameter space. If the closed path had been on a flat surface there would be no anholonomy. The Berry phase is a quantum version of parallel transport, in which the Bloch Hamiltonian plays the role of the curved surface and the wavevectors \textbf{k} act as the parameters that define the closed path. For obvious reasons, the Berry phase is often called a geometrical phase.

\subsection{Berry curvature}\label{subsec:BerryCurv}

\begin{figure}
\vspace*{-2cm}
\setlength{\abovecaptionskip}{-10pt plus 0pt minus 0pt}
\centering
\includegraphics[width=0.5\textwidth, angle=90]{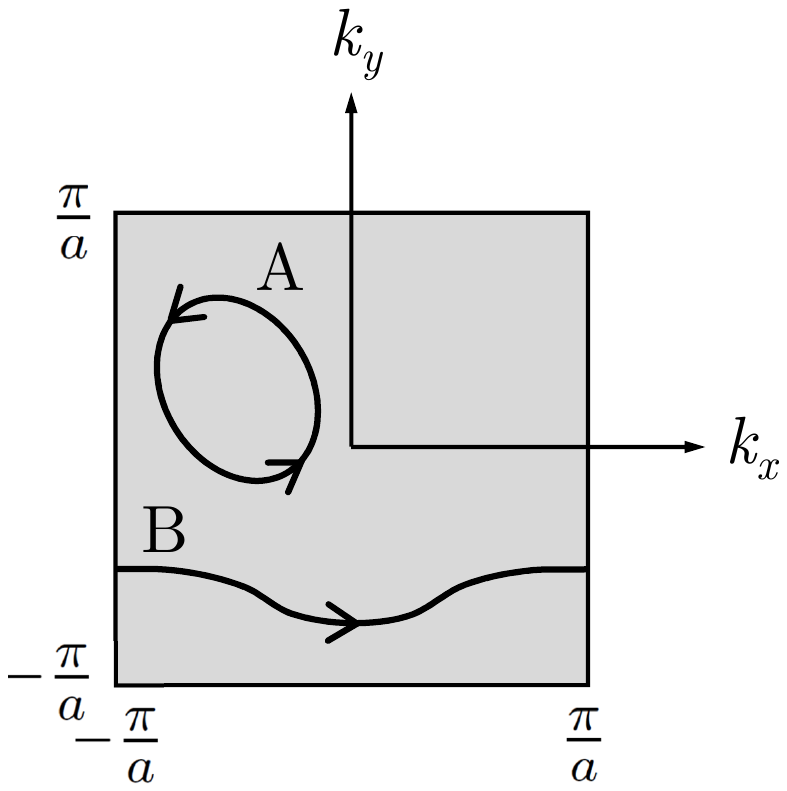}\vspace{0cm}
\caption{Closed paths in two-dimensional \textbf{k} space. The shaded square is the BZ of a square lattice (lattice parameter $a$). Path A is contractible and path B is non-contractible. Path B is still a closed path because points on opposite sides of the BZ related by a reciprocal lattice vector are equivalent.} \label{fig:Paths}
\end{figure}

At this point we need to make a clarification about closed paths in \textbf{k} space. Consider the two paths drawn on the 2D BZ in Figure~\ref{fig:Paths}. The simplest type of closed path is a \emph{contractible path} (path A), i.e.~one that can be continuously shrunk to a point. However, because Bloch functions are periodic in the Brillouin zone a \emph{non-contractible} closed path is also possible (path B), where the path crosses the BZ boundary then wraps around to the equivalent point on the opposite face of the BZ. The Berry phase associated with the latter type of path is often called a Zak phase \cite{Zak1989}.

In the case of a contractible path $\mathcal{C}$, one can use Stokes' theorem to convert the contour integral in Equation~(\ref{eq:Berry-phase-6}) into a surface integral:
\begin{align}
\oint_\mathcal{C} \boldsymbol{\mathcal{A}}_n\cdot\textrm{d}\textbf{k} = \int_\mathcal{S} (\nabla_\textbf{k} \times \boldsymbol{\mathcal{A}}_n)\cdot\textrm{d}\textbf{S},
\label{eq:Berry-Stokes}
\end{align}
where $\mathcal{S}$ is a surface in \textbf{k} space bounded by $\mathcal{C}$. For Equation~(\ref{eq:Berry-Stokes}) to apply, $\boldsymbol{\mathcal{A}}_n(\textbf{k})$ must be well defined over $\mathcal{S}$. Assuming that to be the case, the phase change around $\mathcal{C}$ may be written
\begin{equation}
\gamma_n = \int_\mathcal{S}\boldsymbol{\mathcal{B}}_n \cdot\textrm{d}\textbf{S},
\label{eq:Berry-curvature-1}
\end{equation}
where
\begin{equation}
\boldsymbol{\mathcal{B}}_n = \nabla_\textbf{k} \times \boldsymbol{\mathcal{A}}_n
\label{eq:Berry-curvature-curl}
\end{equation}
is an emergent gauge field called the \emph{Berry curvature}, and is a \textbf{k}-space analogue of the magnetic induction.\footnote{In the literature, the Berry curvature is often denoted by ${\boldsymbol\Omega}$ or \textbf{F}. Here, I use $\boldsymbol{\mathcal{A}}$ and $\boldsymbol{\mathcal{B}}$ for the Berry potential and Berry curvature to reinforce the analogy with magnetic fields.} We see from (\ref{eq:Berry-curvature-1}) that the Berry phase around $\mathcal{C}$ is the flux of Berry curvature (\emph{Berry flux}) through the surface $\mathcal{S}$.

From Equations~(\ref{eq:Berry-connection-gauge-dep}) and (\ref{eq:Berry-curvature-curl}), together with the identity $\nabla \times (\nabla \beta) = 0$, one can see that $\boldsymbol{\mathcal{B}}_n$ is gauge-invariant and therefore an observable quantity. Since $u_n$ and $\psi_n$ are related by a phase factor, Equation~(\ref{eq:Bloch-fn}), and $\psi_n$ is periodic in the reciprocal lattice, it follows that $\boldsymbol{\mathcal{B}}_n$ is also periodic in the reciprocal lattice.

From (\ref{eq:Berry-curvature-1}) it looks like $\gamma_n$ should also be gauge-invariant. As we shall show in Section~\ref{sec:Chern-theorem}, $\gamma_n$ is actually gauge invariant to within an integer multiple of $2\pi$. Hence, the Berry phase factor $\textrm{e}^{-i\gamma_n}$ is observable too. Note that the Zak phase is defined from a non-contractible path, and so it cannot be expressed in term of a Berry curvature.

\subsection{Berry curvature in terms of virtual transitions}\label{subsec:other-bands}

Although the results derived above for $\gamma_n$ and $\boldsymbol{\mathcal{B}}_n$ are expressed purely in terms of Bloch functions for a single band $n$, a non-vanishing Berry phase/curvature actually depends upon the existence of other bands $m \ne n$. To see this, combine Equations~(\ref{eq:Berry-connection}), (\ref{eq:Berry-curvature-1}) and (\ref{eq:Berry-curvature-curl}), and write (omitting the \textbf{k} dependence for simplicity)
\begin{align}
\gamma_n & = -\textrm{Im}\int_\mathcal{S} (\nabla \times \langle u_n|\nabla |u_n\rangle)\cdot\textrm{d}\textbf{S}\nonumber\\
& = -\textrm{Im}\int_\mathcal{S} \langle \nabla u_n|\times|\nabla u_n\rangle \cdot\textrm{d}\textbf{S}\nonumber\\
& = -\textrm{Im}\int_\mathcal{S} \sum_{m\ne n} \langle\nabla u_n|u_m\rangle\times\langle u_m|\nabla u_n\rangle \cdot\textrm{d}\textbf{S}.
\label{eq:Berry-phase-8}
\end{align}
The last line is obtained by the quantum mechanical closure relation, and the $n=m$ term is omitted from the summation because this term is real and we are only concerned with the imaginary part.  An alternative form of Equation~(\ref{eq:Berry-phase-8}) is obtained from the following identity,\footnote{One can obtain Equation~(\ref{nabla-identity}) by applying the $\nabla$ operator to the Schr\"{o}dinger equation (\ref{eq:TISE-Hk}),
\begin{align}
\nabla ({\mathcal H} |u_n\rangle) & = E_n\nabla |u_n\rangle + (\nabla E_n)|u_n\rangle,\nonumber
\end{align}
and then using the fact that $\langle u_m|u_n\rangle = \delta_{mn}$ to show that for $m \ne n$,
\begin{align}
 E_n\langle u_m|\nabla| u_n\rangle & = \langle u_m |\nabla {\mathcal H} |u_n\rangle\nonumber\\
 & =  \langle u_m |{\mathcal H} |\nabla u_n\rangle + \langle u_m |(\nabla {\mathcal H})|u_n\rangle\nonumber\\
 & = E_m\langle u_m|\nabla |u_n\rangle + \langle u_m |(\nabla {\mathcal H})|u_n\rangle,\nonumber
\end{align}
which leads directly to Equation~(\ref{nabla-identity}).}
\begin{equation}
\langle u_m|\nabla | u_n\rangle = \frac{\langle u_m |(\nabla {\mathcal H})|u_n\rangle}{E_n-E_m}, \hspace{30pt}n \ne m.
\label{nabla-identity}
\end{equation}
Putting (\ref{nabla-identity}) in (\ref{eq:Berry-phase-8}), and comparing with (\ref{eq:Berry-curvature-1}), one obtains
\begin{equation}
\boldsymbol{\mathcal{B}}_n = -\textrm{Im}\sum_{m\ne n}\frac{\langle u_n |(\nabla {\mathcal H})|u_m\rangle \times \langle u_m |(\nabla {\mathcal H})|u_n\rangle}{(E_n-E_m)^2}.
\label{eq:Berry-curvature-2}
\end{equation}
Formula (\ref{eq:Berry-curvature-2}) shows that the Berry curvature in the $n^\textrm{th}$ band is determined by virtual transitions to other bands, and that bands which are closest in energy to band $n$ will tend to contribute the most. In other words, the Berry curvature is an effective (emergent) field which arises when we project the effects of inter-band transitions onto the motion of an electronic quasiparticle in a single band.

Finally, note that $\sum_n \boldsymbol{\mathcal{B}}_n = 0$, and hence $\sum_n \gamma_n = 0$. This follows because the numerator in (\ref{eq:Berry-curvature-2}) has the opposite sign for the terms $(n,m)$ and $(m,n)$ in the double summation, so these terms cancel.

\subsection{Linear band crossings in 3D}\label{subsec:2-level}

\begin{figure}
\vspace*{-2.5cm}
\setlength{\abovecaptionskip}{-20pt plus 0pt minus 0pt}
\centering
\includegraphics[width=0.6\textwidth, angle=90]{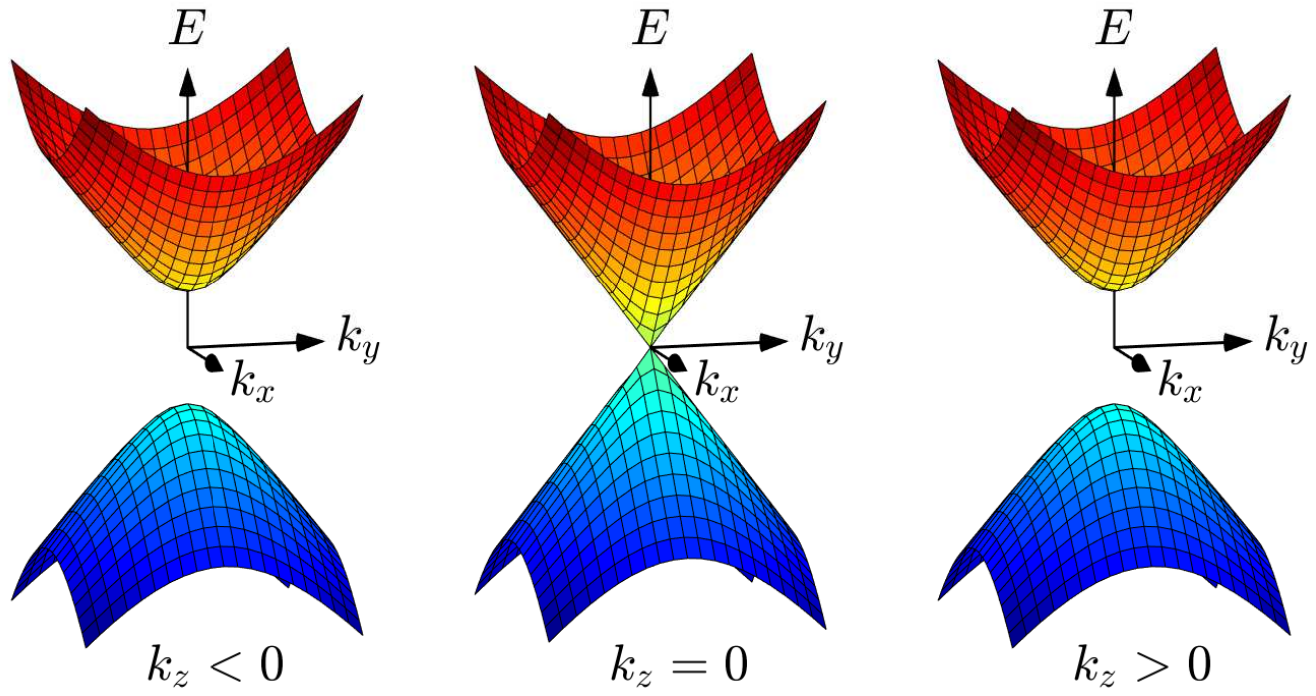}\vspace{0cm}
\caption{Dispersion surfaces in the vicinity of a linear band crossing point at $\textbf{k} = 0$. The upper and lower bands $E_\pm(\textbf{k})$, calculated from Equation~(\ref{eq:H(k)-2}), are plotted as a function of $(k_x,k_y)$, for three $k_z$ values that span the crossing point.} \label{fig:WeylEk}
\end{figure}

Suppose that the contour $\mathcal{C}$ lies close to a point $P$ where two or more bands cross.\footnote{The degeneracy point $P$ must not lie on $\mathcal{C}$, otherwise the adiabatic condition is not satisfied, see Section~\ref{subsec:adiabatic}.}  The denominator of Equation~(\ref{eq:Berry-curvature-2}) implies that the Berry curvature for band $n$ is dominated by the contributions from the other level(s) $m$ involved in the degeneracy. For simplicity, we shall consider a two-level system, and without loss of generality we take P to be at $(E,\textbf{k}) = (0,0)$. We shall also assume for now that the dispersion $E_n(\textbf{k})$ is linear and isotropic (the case of an anisotropic dispersion is discussed below).

As there are two Bloch functions for each \textbf{k} the Hamiltonian can be represented by a $2 \times 2$ Hermitian matrix, which in the vicinity of P takes the form
\begin{align}
{\mathcal H}(\textbf{k}) & = \hbar v\left(\begin{array}{c c} k_z & k_x-ik_y\\[2pt] k_x+ik_y & -k_z \end{array} \right).
\label{eq:H(k)-1}
\end{align}
We see that the parameters of $\mathcal{H}$ are $k_x, k_y$ and $k_z$. The eigenvalues of (\ref{eq:H(k)-1}) are
\begin{align}
E_\pm(\textbf{k})  & = \pm\hbar v \sqrt{k_x^2+k_y^2+k_z^2},\nonumber\\
& = \pm\hbar v k,
\label{eq:H(k)-2}
\end{align}
which is the assumed linear dispersion with an isotropic speed $v$, see Figure~\ref{fig:WeylEk}. The normalised eigenvectors written in spherical coordinates in \textbf{k} space\footnote{$(k_x, k_y, k_z) = k (\sin \theta_\textbf{k} \cos \phi_\textbf{k},\,\sin \theta_\textbf{k} \sin\phi_\textbf{k},\, \cos\theta_\textbf{k})$.} are
\begin{align}
|+\rangle = \left(\begin{array}{c}\cos \frac{\theta_\textbf{k}}{2}\,\textrm{e}^{-i\phi_\textbf{k}} \\[2pt] \sin\frac{\theta_\textbf{k}}{2} \end{array}\right), \hspace{20pt} |-\rangle = \left(\begin{array}{c}\sin\frac{\theta_\textbf{k}}{2}\,\textrm{e}^{-i\phi_\textbf{k}} \\[2pt] -\cos\frac{\theta_\textbf{k}}{2} \end{array}\right).
\label{eq:H(k)-3}
\end{align}
The two components of the eigenvectors represent the amplitudes of the two basis vectors of the two-level system, e.g.~two different orbitals.

We can now calculate the Berry curvature. In spherical coordinates $\nabla_\textbf{k} = \hat{\textbf{k}}(\frac{\partial}{\partial k}) + \hat{{\boldsymbol\theta}}_\textbf{k}(\frac{1}{k}\frac{\partial}{\partial \theta_\textbf{k}}) + \hat{{\boldsymbol\phi}}_\textbf{k}(\frac{1}{k\sin\theta_\textbf{k}}\frac{\partial}{\partial \phi_\textbf{k}})$, and hence from Equation~(\ref{eq:Berry-connection}) we find
\begin{align}
\boldsymbol{\mathcal{A}}_+(\textbf{k}) &  = \frac{1}{2k}\cot \mbox{$\frac{\theta_\textbf{k}}{2}$}\, \hat{{\boldsymbol\phi}}_\textbf{k}, \hspace{20pt}\boldsymbol{\mathcal{A}}_-(\textbf{k}) = \frac{1}{2k}\tan \mbox{$\frac{\theta_\textbf{k}}{2}$}\, \hat{{\boldsymbol\phi}}_\textbf{k},
\label{eq:Berry-connection-2-level}
\end{align}
for the $|+\rangle$ and $|-\rangle$ eigenstates. By writing the curl of a vector in spherical coordinates\footnote{
\begin{align}
\nabla_\textbf{k} \times \textbf{F} = & \frac{1}{k\sin \theta_\textbf{k}}\left\{\frac{\partial}{\partial \theta_\textbf{k}}(F_\phi \sin \theta_\textbf{k}) -  \frac{\partial F_\theta}{\partial \phi_\textbf{k}}\right\}\hat{\textbf{k}}\nonumber\\
& + \frac{1}{k}\left\{\frac{1}{\sin \theta_\textbf{k}} \frac{\partial F_k}{\partial \phi_\textbf{k}} - \frac{\partial}{\partial k}(k F_\phi)\right\}\hat{{\boldsymbol\theta}}_\textbf{k}\nonumber\\ & + \frac{1}{k}\left\{\frac{\partial}{\partial k}(k F_\theta) - \frac{\partial F_k}{\partial \theta}\right\}\hat{{\boldsymbol\phi}}_\textbf{k},\nonumber
\end{align}
} we can obtain from Equations~(\ref{eq:Berry-curvature-curl}) and (\ref{eq:Berry-connection-2-level})
\begin{align}
\boldsymbol{\mathcal{B}}_\pm(\textbf{k}) = \mp \frac{1}{2k^2}\hat{\textbf{k}}.
\label{eq:Berry-curvature-2-level}
\end{align}

\begin{figure}
\vspace*{-0.5cm}
\setlength{\abovecaptionskip}{-10pt plus 0pt minus 0pt}
\centering
\includegraphics[width=0.5\textwidth, angle=0]{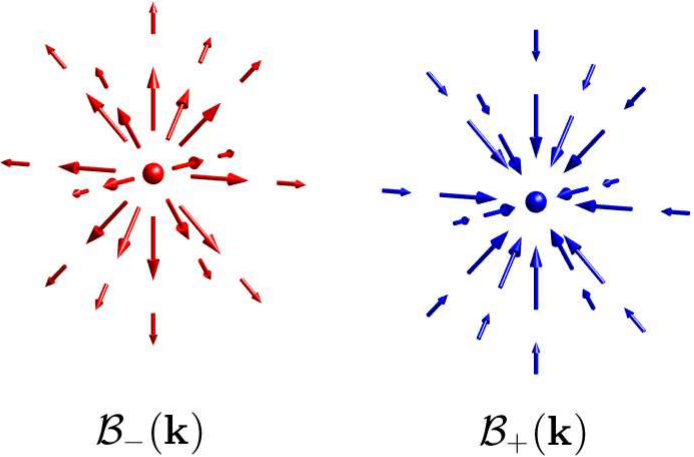}\vspace{1.5cm}
\caption{Representation of the Berry curvature associated with the lower (left) and upper (right) bands that form a Weyl point, Equation~(\ref{eq:Berry-curvature-2-level}). Both vector fields, $\boldsymbol{\mathcal{B}}_-(\textbf{k})$ and $\boldsymbol{\mathcal{B}}_+(\textbf{k})$, are centred on $\textbf{k}=0$.} \label{fig:BerryCurv}
\end{figure}

We see that $\boldsymbol{\mathcal{B}}_-(\textbf{k})$ and $\boldsymbol{\mathcal{B}}_+(\textbf{k})$ resemble the fields from point charges (monopoles) in \textbf{k} space of strength $+\frac{1}{2}$ and $-\frac{1}{2}$, respectively, see Figure~\ref{fig:BerryCurv}.  The positive monopole charge is associated with the negative energy solution, and \textit{vice versa}. Although the oppositely charged monopoles coincide at the nodes, the associated Berry curvature fields do not cancel one another because they belong to different bands. Degeneracy points, therefore, are sources and sinks of Berry curvature.

The Berry phase for adiabatic transport around $\mathcal{C}$ can be obtained from either the Berry connection, Equations~(\ref{eq:Berry-phase-6}) and (\ref{eq:Berry-connection-2-level}), or the Berry curvature, Equations~(\ref{eq:Berry-curvature-1}) and (\ref{eq:Berry-curvature-2-level}). The result will be the same, modulo $2\pi$. In the latter case, we have
\begin{align}
\gamma_\pm & = \mp\frac{1}{2}\int_\mathcal{S}\frac{\hat{\textbf{k}}\cdot\textrm{d}\textbf{S}}{k^2}\nonumber\\
& = \mp\frac{\Omega}{2},
\label{eq:Berry-phase-linear}
\end{align}
where $\Omega$ is the solid angle subtended by $\mathcal{S}$ at $P$. A special case is when $\mathcal{C}$ is restricted to a plane which contains $P$. Then, $\gamma_\pm = \mp \pi$ if $\mathcal{C}$ encloses $P$, or $\gamma_\pm = 0$ otherwise.

The crossing point of two singly degenerate bands in 3D \textbf{k} space is called a \emph{Weyl point} or \emph{Weyl node}, for the reason given in Section~\ref{sec:Dirac-Weyl}. It is conventional to assign a chirality $\chi$ to the nodes, such that $\chi = +1$ if the lower-energy band acts as a source of Berry curvature (positive monopole charge) and $\chi = -1$ if the lower-energy band acts as a sink of Berry curvature (negative monopole charge). The Weyl node described above has $\chi = +1$. A change of sign in front of the Hamiltonian~(\ref{eq:H(k)-1}) generates the opposite chirality. Electronic quasiparticles in the vicinity of Weyl nodes have special properties which differ from those of conventional electrons. We will elaborate more in Section~\ref{sec:Dirac-Weyl}.

\subsection{Chern theorem and Chern number}\label{sec:Chern-theorem}

From Equation~(\ref{eq:Berry-curvature-2-level}) and Gauss' theorem, it follows that the integral of $\boldsymbol{\mathcal{B}}_\pm(\textbf{k})$ over a surface surrounding P in \textbf{k} space is $\mp 2\pi$.  This is a particular case of the \textit{Chern theorem}, which states that the Berry flux calculated for any closed 2D surface $\mathcal{S}$ in \textbf{k} space is quantised in multiples of $2\pi$, i.e.
\begin{align}
\oint_\mathcal{S} \boldsymbol{\mathcal{B}}_n \cdot \textrm{d}\textbf{S} = 2\pi C_n.
\label{eq:Chern_thm}
\end{align}
The integer $C_n$ is a topological invariant known as the \emph{Chern number}. The Fermi surface can be such a closed surface, in which case the Fermi surface will have a well-defined Chern number. For the isotropic band crossing analysed in Section~\ref{subsec:2-level} the Fermi surface is spherical, and depending whether the Fermi level is higher or lower in energy than the Weyl node (i.e.~whether the Fermi surface is electron-like or hole-like), the Fermi surface has a Chern number $C_n = -1$ or $+1$, respectively.

\subsection*{Proof of the Chern theorem}\label{sec:Chern-theorem-proof}

To understand the Chern theorem (\ref{eq:Chern_thm}) better we need to appreciate the effect of a gauge transformation on the different expressions for the Berry phase, specifically (see Sections~\ref{subsec:BerryPhaseConn} and \ref{subsec:BerryCurv})
\begin{align}
\gamma = \oint_\mathcal{C} \boldsymbol{\mathcal{A}}(\textbf{k})\cdot \textrm{d}\textbf{k} = \int_\mathcal{S}\boldsymbol{\mathcal{B}} \cdot\textrm{d}\textbf{S}.
\label{eq:Berry-phase-7}
\end{align}
The band index $n$ is suppressed here to simplify the notation. Now consider the gauge transformation
\begin{align}
|u'(\textbf{k})\rangle & =\textrm{e}^{-i\beta(\textbf{k})}|u(\textbf{k})\rangle.
\label{eq:uk-gauge-transfn}
\end{align}
If the functions $u(\textbf{k})$ and $u'(\textbf{k})$ are to be well-behaved they must be single-valued functions of \textbf{k}, and hence $\exp(-i\beta)$ must also be single-valued.  Therefore, around any closed path $\mathcal{C}$ in \textbf{k} space the initial and final values of $\beta$ must satisfy
\begin{align}
\beta_\textrm{f}-\beta_\textrm{i} = 2\pi m,\ \ \ \ \ (m = \textrm{integer}).
\label{eq:beta-change}
\end{align}
The effect of a gauge transformation on $\boldsymbol{\mathcal{A}}$ was given in Equation~(\ref{eq:Berry-connection-gauge-dep}),
\begin{align}
\boldsymbol{\mathcal{A}}' & = \boldsymbol{\mathcal{A}} + \nabla_\textbf{k} \beta(\textbf{k}).
\label{eq:Berry-connection-gauge-dep-2}
\end{align}
Integrating this equation around $\mathcal{C}$ and using (\ref{eq:Berry-phase-6}) and (\ref{eq:beta-change}), we obtain
\begin{align}
\gamma' = \gamma + 2\pi m.
\label{eq:gamma-gauge-transfn}
\end{align}
At first glance this result seems at odds with Equation~(\ref{eq:Berry-phase-7}). Previously we found that the Berry curvature  $\boldsymbol{\mathcal{B}}$ is gauge-invariant, which according to (\ref{eq:Berry-phase-7}) implies that $\gamma$ is gauge-invariant, whereas Equation~(\ref{eq:gamma-gauge-transfn}) shows that under a gauge transformation $\gamma$ can change by an integer multiple of $2\pi$. The reason for this apparent paradox is that the second equality in (\ref{eq:Berry-phase-7}) came from the application of Stokes' theorem, which requires $\boldsymbol{\mathcal{A}}$ to be well-behaved over the entire surface $\mathcal{S}$. If we choose a gauge in which this requirement is satisfied, then Equation~(\ref{eq:Berry-phase-7}) will hold as an equality. If not, (\ref{eq:Berry-phase-7}) will only hold modulo $2\pi$.

\begin{figure}
\vspace*{-4cm}
\setlength{\abovecaptionskip}{-80pt plus 0pt minus 0pt}
\centering
\includegraphics[width=0.6\textwidth, angle=0]{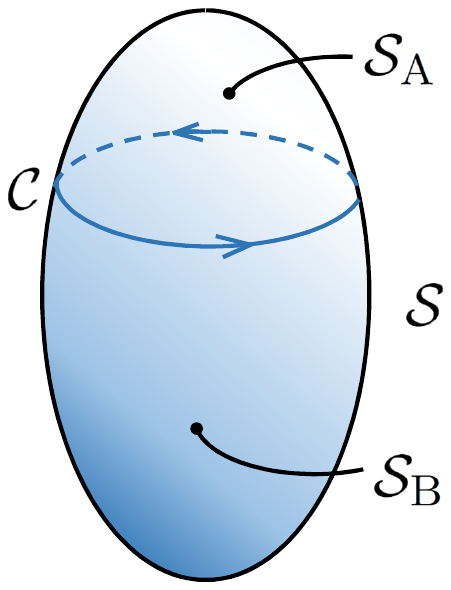}\vspace{0cm}
\caption{Surfaces for proof of the Chern theorem. The closed surface $\mathcal{S}$ is divided into two surfaces $\mathcal{S}_\textrm{A}$ and $\mathcal{S}_\textrm{B}$ having a common boundary $\mathcal{C}$. } \label{fig:Chern}
\end{figure}
In order to prove the Chern theorem, we shall evaluate the integral of $\boldsymbol{\mathcal{B}}$ over a closed 2D surface $\mathcal{S}$, using what we have just learnt about the effects of a gauge transformation. To proceed, we divide $\mathcal{S}$ into two regions A and B having a common boundary $\mathcal{C}$, so that $\mathcal{S} = \mathcal{S}_\textrm{A} + \mathcal{S}_\textrm{B}$, see Figure~\ref{fig:Chern}. We choose a gauge such that Stokes' law applies in region A, and calculate the Berry phase $\gamma_\textrm{A}$ from the first equality in (\ref{eq:Berry-phase-7}). The line integral around $\mathcal{C}$ is taken in the sense related by the right-hand rule to the outward normal of $\mathcal{S}_\textrm{A}$. We then repeat for region B, but choosing a gauge in which Stokes' law applies in B. The gauges used for B and A need not be the same. The direction of circulation around $\mathcal{C}$ is reversed because the outward normal of B is opposite to that of A, so the line integral gives $-\gamma_\textrm{B}$. Combining the two terms, we obtain
\begin{align}
\oint_{\mathcal{S}}\boldsymbol{\mathcal{B}} \cdot\textrm{d}\textbf{S} & = \int_{\mathcal{S}_\textrm{A}}\boldsymbol{\mathcal{B}} \cdot\textrm{d}\textbf{S} + \int_{\mathcal{S}_\textrm{B}}\boldsymbol{\mathcal{B}} \cdot\textrm{d}\textbf{S}\nonumber\\ & =\gamma_\textrm{A} - \gamma_\textrm{B}.
\label{eq:Chern-theorem-AB}
\end{align}
If the gauges used for regions A and B are the same then $\gamma_\textrm{A} = \gamma_\textrm{B}$ and the total Berry flux is zero. However, if the gauges are different, Equation~(\ref{eq:gamma-gauge-transfn}) tells us that $\gamma_\textrm{A}$ and $\gamma_\textrm{B}$ can differ by an integer multiple of $2\pi$. Hence, Equation~(\ref{eq:Chern-theorem-AB}) reduces to Equation~(\ref{eq:Chern_thm}), which proves the Chern theorem.

The existence of a non-zero Chern number signals that $\boldsymbol{\mathcal{A}}$ is not well-behaved over the entirety of $\mathcal{S}$. In other words, there is some kind of `twist' in the Bloch function manifold. For example, in the band-crossing model presented in Section~\ref{subsec:2-level} one can see from Equation~(\ref{eq:Berry-connection-2-level}) that there are singularities in $\boldsymbol{\mathcal{A}}_\pm$ at the poles, $\theta = 0$ or $\pi$.  Therefore, a non-zero Chern number tells us that the Bloch function manifold has a non-trivial topology on $\mathcal{S}$.

\subsection{Fermion doubling theorem}\label{sec:Fermion-doubling-theorem}

The Chern theorem leads us to another fundamental result known as the Nielsen--Ninomiya \textit{fermion doubling} theorem~\cite{NielsenNinomiya1983}, which requires that the sum of the chiralities of the Weyl nodes in a 3D BZ must vanish,
\begin{align}
\sum_{j \in \text{BZ}} \chi_j = 0.
\label{Nielsen-Ninomiya-thm}
\end{align}
It follows that Weyl nodes always come in pairs with opposite chirality (hence, \emph{fermion doubling}).

To prove this result, let us choose $\mathcal{S}$ to enclose the entire BZ, i.e.~$\mathcal{S}$ is formed by the BZ boundary. Recall that the Berry curvature is periodic in the reciprocal lattice. On opposite faces of the BZ, therefore, the Berry curvature is the same but the outward normals of the faces point in opposite directions. Hence, the contribution to the integral in Equation~(\ref{eq:Chern_thm}) from such a pair of BZ faces cancels. For any band, therefore, the net Chern number defined over the entire BZ surface must vanish. Equation~(\ref{Nielsen-Ninomiya-thm}) then follows immediately because the Chern number of a closed surface is simply the sum of the chiralities of the Weyl nodes inside the surface.

\section{Symmetry considerations}\label{sec:symmetry}

Symmetry plays a key r\^{o}le in determining the existence and distribution of band degeneracies in \textbf{k} space.

\subsection{Symmetry and degeneracy}\label{subsec:symmetry-deg}

If a system possesses a symmetry $S$ it means that the system is invariant under $S$, i.e.~it looks the same before and after the $S$ transformation. Another way to say this is that the energy of the system remains unchanged under $S$.

To express this idea and its consequences in the language of quantum mechanics, let $U_s$ be the quantum mechanical operator that represents $S$, and let $|\psi\rangle$ be an eigenstate of the system Hamiltonian $\mathcal{H}$ with eigenvalue $E$,
\begin{align}
\mathcal{H}|\psi\rangle = E|\psi\rangle.
\label{eq:HU-commute-1}
\end{align}
If the state $U_s|\psi\rangle$ is to have the same energy as $|\psi\rangle$, then
\begin{align}
\mathcal{H}U_s|\psi\rangle & = EU_s |\psi\rangle\nonumber\\
& = U_s\mathcal{H}|\psi\rangle.
\label{eq:HU-commute-2}
\end{align}
Therefore, we conclude that a system is invariant under $S$ if $U_s$ commutes with $\mathcal{H}$.

Now let us reverse the argument. Suppose there is a symmetry operator $U_s$ that commutes with $\mathcal{H}$. It is easy to see from (\ref{eq:HU-commute-2}) that if $|\psi\rangle$ is an eigenstate of $\mathcal{H}$  then $U_s|\psi\rangle = |\phi\rangle$ is also an eigenstate of $\mathcal{H}$ with the same energy eigenvalue $E$. Hence, if $|\psi\rangle$ and $|\phi\rangle$ are distinct (i.e.~linearly independent) eigenstates, then the energy level $E$ is degenerate. In other words, the $S$ symmetry of $\mathcal{H}$ enforces a degeneracy.

This analysis illustrates how symmetry can cause degeneracy. Conversely, when one observes degenerate eigenstates there is often (but not always) an underlying symmetry that is responsible for the degeneracy. When degeneracy is associated with symmetry it is called \emph{symmetry-protected degeneracy}. When there is no symmetry involved, it is called \emph{accidental degeneracy}.

\subsection{Inversion and time-reversal symmetry}\label{subsec:symmetry-PT}

The inversion (or parity) transformation $\mathcal{P}$ maps $\textbf{r}$ onto $-\textbf{r}$, relative to a suitably defined origin. If we apply $\mathcal{P}$ to a Bloch function  $\psi_\textbf{k}(\textbf{r})$ (see Section~\ref{subsec:Bloch-fns}) we obtain another Bloch function $\psi_{-\textbf{k}}(\textbf{r})$:
\begin{align}
\mathcal{P}\psi_\textbf{k}(\textbf{r}) & = \textrm{e}^{-i\textbf{k}\cdot\textbf{r}}u_\textbf{k}(-\textbf{r})\nonumber\\
& = \textrm{e}^{i(-\textbf{k})\cdot\textbf{r}}u_{-\textbf{k}}(\textbf{r})\nonumber\\
& = \psi_{-\textbf{k}}(\textbf{r})
\label{eq:Bloch-inversion}
\end{align}
Here we have simply changed the name of $u_\textbf{k}(-\textbf{r})$ to $u_{-\textbf{k}}(\textbf{r})$. The function $\psi_{-\textbf{k}}(\textbf{r})$ is a Bloch function because $u_{-\textbf{k}}(\textbf{r})$ is cell-periodic: $u_{-\textbf{k}}(\textbf{r}+\textbf{R})=u_{\textbf{k}}(-\textbf{r}-\textbf{R})=u_{\textbf{k}}(-\textbf{r})=u_{-\textbf{k}}(\textbf{r})$. Note that all lattices have inversion symmetry, so if $\textbf{R}$ is a lattice vector then $-\textbf{R}$ is one too. We shall sometimes write that $\mathcal{P}$ maps $\textbf{k}$ to $-\textbf{k}$, by which we mean is that $\mathcal{P}$ transforms the Bloch function associated with $\textbf{k}$ into the Bloch function associated with $-\textbf{k}$.

The treatment of time-reversal symmetry is more subtle. In classical mechanics, the equation of motion for a conservative force is invariant under $t \rightarrow -t$, and we expect a similar relation to hold in quantum mechanics. However, a simple reversal of time in Equation~(\ref{eq:TDSE}) does not produce the desired effect. Instead, one needs to define an operator $\mathcal{T}$ which performs complex conjugation in addition to reversing time. Such an operator is termed \emph{antiunitary}, and has the antilinear property,
\begin{align}
\mathcal{T}c|\psi\rangle & = c^\ast\mathcal{T}|\psi\rangle,
\label{eq:antiunitary}
\end{align}
for a complex number $c$. In contrast, $\mathcal{P}$ is a \emph{unitary} operator, $\mathcal{P}c|\psi\rangle = c\mathcal{P}|\psi\rangle$.

As a simple example, consider the effect of $\mathcal{T}$ on the function $\psi = \textrm{e}^{i(kx-\omega t)}$, which describes a plane wave travelling in the $+x$ direction. Reversing the sign of $t$ and taking the complex conjugate we obtain $\mathcal{T}\psi = \textrm{e}^{i(-kx-\omega t)}$, which is a plane wave travelling in the $-x$ direction, as expected. We also see that in this example, $\mathcal{T}$ is equivalent to changing $k$ into $-k$.  Next, apply $\mathcal{T}$ to a Bloch function:
\begin{align}
\mathcal{T}\psi_{\textbf{k}\uparrow}(\textbf{r}) & = \textrm{e}^{-i\textbf{k}\cdot\textbf{r}}u_{\textbf{k}\downarrow}^\ast(\textbf{r})\nonumber\\
& = \textrm{e}^{i(-\textbf{k})\cdot\textbf{r}}u_{-\textbf{k}\downarrow}(\textbf{r})\nonumber\\
& = \psi_{-\textbf{k}\downarrow}(\textbf{r}).
\label{eq:Bloch-TR}
\end{align}
We have added an arrow subscript to indicate any spin-like degrees of freedom in the Bloch function (in general, $\uparrow$ can represent any dynamical variable that transforms like spin). This addition is needed because $\mathcal{T}$ reverses the direction of spin. $\mathcal{P}$, on the other hand, does not reverse the direction of spin, so if spin were included in Equation~(\ref{eq:Bloch-inversion}) it would be written $\mathcal{P}\psi_{\textbf{k}\uparrow}(\textbf{r}) = \psi_{-\textbf{k}\uparrow}(\textbf{r})$. We have renamed $u_{\textbf{k}\downarrow}^\ast(\textbf{r})$ as $u_{-\textbf{k}\downarrow}(\textbf{r})$ in Equation~(\ref{eq:Bloch-TR}), very much like in Equation~(\ref{eq:Bloch-inversion}).  Overall, we see that the effect of $\mathcal{T}$ is equivalent to changing a Bloch function associated with $\textbf{k}$ into one with $-\textbf{k}$ and the opposite spin.

Time-reversal symmetry breaking in crystalline solids is usually associated with magnetic order or the presence of magnetic fields, because magnetic moments and fields are generated by circulating currents, and currents reverse direction when time is reversed.  The presence or absence of inversion symmetry can be identified from the crystallographic space group or, if the material is magnetically ordered, from the magnetic space group.

\subsection{Kramers' theorem}\label{subsec:Kramers}

Further analysis of the time-reversal symmetry operation in quantum mechanics \cite{Schiff1968} reveals that for a state $|\psi\rangle$ with total angular momentum quantum number $J$,
\begin{align}
\mathcal{T}^2|\psi\rangle = (-1)^{2J}|\psi\rangle.
\label{eq:Kramers-1}
\end{align}
We can use this result to prove a general property of half-integer spin systems with time-reversal symmetry, which is that every energy level is $n$-fold degenerate, where $n$ is an even integer (not necessarily the same for each level). This is known as Kramers' theorem.

We infer from the arguments in Section~\ref{subsec:symmetry-deg} that if $|\psi\rangle$ is an eigenfunction of a $\mathcal{T}$-invariant Hamiltonian, then $|\psi\rangle$ and $\mathcal{T}|\psi\rangle$ are degenerate eigenfunctions providing they are not linearly related. Let us assume the opposite, that $\mathcal{T}|\psi\rangle = c|\psi\rangle$. In that case, the antilinear property (\ref{eq:antiunitary}) means that $\mathcal{T}^2|\psi\rangle = \mathcal{T}c|\psi\rangle = c^\ast \mathcal{T}|\psi\rangle = |c|^2|\psi\rangle$. Since $|c|^2 > 0$, our assumption contradicts Equation~(\ref{eq:Kramers-1}), which says that for a half-integer spin system $\mathcal{T}^2|\psi\rangle = -|\psi\rangle$. We conclude, therefore, that $|\psi\rangle$ and $\mathcal{T}|\psi\rangle$ must be distinct eigenfunctions. It follows that the eigenfunctions belonging to each energy level come in degenerate pairs related by time reversal, which establishes the theorem.

\subsection{Effect of $\textbf{P}$ and $\mathcal{T}$ on electronic bands}\label{subsec:symmetry-bands}

\begin{figure}
\vspace*{-3cm}
\setlength{\abovecaptionskip}{-70pt plus 0pt minus 0pt}
\centering
\includegraphics[width=0.5\textwidth, angle=0]{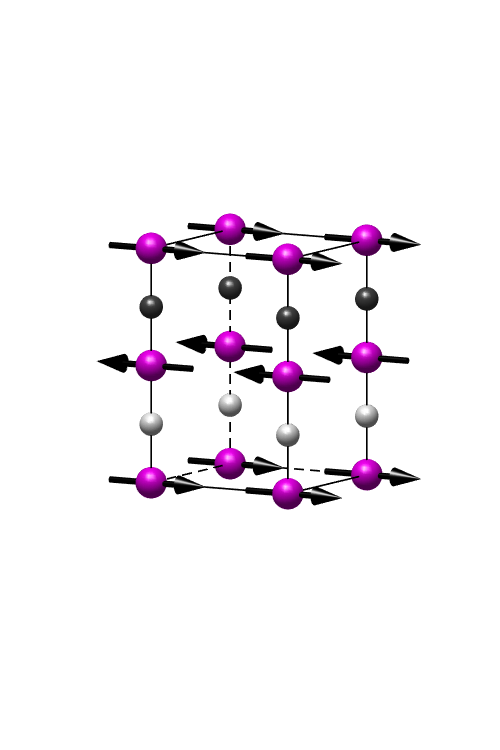}\vspace{0cm}
\caption{An A-type antiferromagnet that has neither $\mathcal{P}$ nor $\mathcal{T}$ symmetry, but has $\mathcal{P}\times \mathcal{T}$ symmetry. In the paramagnetic phase there are centres of inversion symmetry in the layers of light grey and black atoms, at positions $0,0,\frac{1}{4}$ and $0,0,\frac{3}{4}$, respectively. These inversion centres are lost in the antiferromagnetic phase because $\mathcal{P}$ acting through one of these centres takes a spin pointing to the right onto a site where the spin points to the left. Note that the direction of a magnetic moment remains unchanged under $\mathcal{P}$ but reverses under $\mathcal{T}$.} \label{fig:AAFM}
\end{figure}

If a crystal potential has inversion symmetry, then from Equation~(\ref{eq:Bloch-inversion}) the energy of the Bloch states $\textbf{k}$ and $-\textbf{k}$ are the same: $E_n(\textbf{k}) = E_n(-\textbf{k})$. The relation $E_n(\textbf{k}) = E_n(-\textbf{k})$ also holds if the crystal has time-reversal symmetry, even if it does not have inversion symmetry. In this case Equation~(\ref{eq:Bloch-TR}) applies, and the degeneracy is a result of Kramers' theorem.

If both $\mathcal{P}$ and $\mathcal{T}$ are present, then $\mathcal{P}$ combined with $\mathcal{T}$ (written $\mathcal{P} \times \mathcal{T}$) maps $\psi_{\textbf{k}\uparrow}(\textbf{r})\rightarrow \psi_{\textbf{k}\downarrow}(\textbf{r})$, and so Kramers' theorem requires that each band is at least doubly degenerate for all \textbf{k} in the BZ. The same applies for systems which are invariant under $\mathcal{P} \times \mathcal{T}$ but not under $\mathcal{P}$ and $\mathcal{T}$ separately. An example of the latter type of system is the antiferromagnetic structure illustrated in Figure~\ref{fig:AAFM}.

\begin{figure}
\vspace*{0.0cm}
\setlength{\abovecaptionskip}{0pt plus 0pt minus 0pt}
\centering
\includegraphics[width=0.6\textwidth, angle=0]{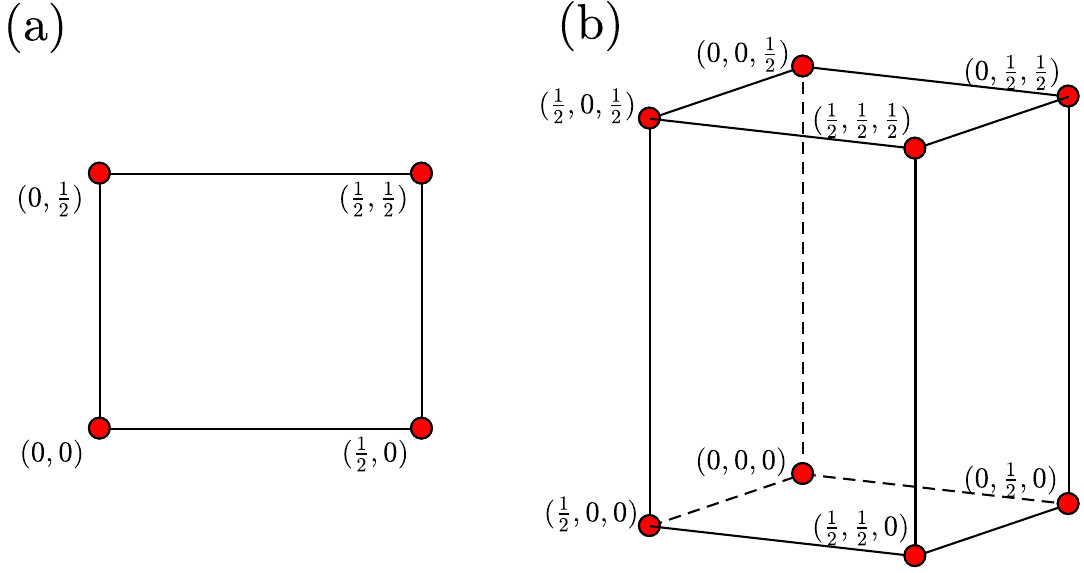}\vspace{1.2cm}
\caption{Time-reversal invariant momenta (TRIM) for (a) two-dimensional, and (b) three-dimensional lattices. The TRIM are shown as red circles, and are the points in reciprocal space where $-\textbf{k} = \textbf{k} + \textbf{G}$. Their coordinates are given in reciprocal lattice units, i.e.~$(h,k,l)$ stands for $h\textbf{b}_1 + k\textbf{b}_2 + l\textbf{b}_3$. In a system with time-reversal symmetry, all states occur in degenerate pairs at the TRIM.} \label{fig:TRIM}
\end{figure}

If only $\mathcal{T}$ or only $\mathcal{P}$ is present, then bands are singly degenerate except at points where bands cross accidentally and at certain special points in the BZ where other spatial or magnetic symmetries leave $\textbf{k}$ invariant, enforcing degeneracies at those positions. An important exception is a degeneracy found in $\mathcal{T}$-invariant systems at special \textbf{k} points called \emph{time-reversal invariant momenta} (TRIM). The TRIM are points where $-\textbf{k} = \textbf{k} + \textbf{G}$ so that $\textbf{k}$ and $-\textbf{k}$ are equivalent. At these points $\mathcal{T}\psi_{\textbf{k}\uparrow}(\textbf{r}) = \psi_{\textbf{k}\downarrow}(\textbf{r})$, and so Kramers degeneracy is enforced by $\mathcal{T}$ even when $\mathcal{P}$ is absent. In $D$ dimensions the number of TRIM is given by $2^D$. Their locations are illustrated in Figure~\ref{fig:TRIM} for $D = 2$ and 3.

\subsection{Symmetry of Berry curvature under $\textbf{P}$ and $\mathcal{T}$}\label{subsec:symmetry-Berry}

As shown in Section~\ref{subsec:symmetry-PT}, the cell-periodic Bloch functions transform under $\mathcal{P}$ and $\mathcal{T}$ according to $\mathcal{P}u_{\textbf{k}\uparrow}(\textbf{r}) = u_{-\textbf{k}\uparrow}(\textbf{r})$ and $\mathcal{T}u_{\textbf{k}\uparrow}(\textbf{r}) = u_{-\textbf{k}\downarrow}(\textbf{r})$. By using these relations in Equations~(\ref{eq:Berry-connection}) and (\ref{eq:Berry-curvature-curl}) one can show that, for a band $n$,
\begin{align}
\mathcal{P}\boldsymbol{\mathcal{B}}_{n}(\textbf{k}) & = \boldsymbol{\mathcal{B}}_{n}(-\textbf{k})\label{eq:Berry-curvature-P}\\
\mathcal{T}\boldsymbol{\mathcal{B}}_{n}(\textbf{k}) & = -\boldsymbol{\mathcal{B}}_{\bar{n}}(-\textbf{k}).
\label{eq:Berry-curvature-T}
\end{align}
Here, $\bar{n}$ stands for the time-reversed partner of $n$. For a crystal with inversion symmetry we have that $\mathcal{P}\boldsymbol{\mathcal{B}}_{n}(\textbf{k}) = \boldsymbol{\mathcal{B}}_{n}(\textbf{k})$, and so Equation~(\ref{eq:Berry-curvature-P}) implies that $\boldsymbol{\mathcal{B}}_{n}(\textbf{k}) = \boldsymbol{\mathcal{B}}_{n}(-\textbf{k})$. Similarly, for $\mathcal{T}$ symmetry, $\mathcal{T}\boldsymbol{\mathcal{B}}_{n}(\textbf{k}) = \boldsymbol{\mathcal{B}}_{\bar{n}}(\textbf{k})$, so Equation~(\ref{eq:Berry-curvature-T}) implies $\boldsymbol{\mathcal{B}}_{n}(\textbf{k}) = -\boldsymbol{\mathcal{B}}_{n}(-\textbf{k})$. In the latter case, the Berry curvature is an odd function of \textbf{k}, which means that the integral of $\boldsymbol{\mathcal{B}}_n(\textbf{k})$ over a BZ must vanish.  Equations (\ref{eq:Berry-curvature-P}) and (\ref{eq:Berry-curvature-T}) also show that the combined operation $\mathcal{P}\times\mathcal{T}$  maps $\boldsymbol{\mathcal{B}}_n(\textbf{k})$ to $-\boldsymbol{\mathcal{B}}_{\bar{n}}(\textbf{k})$, so if the crystal has $\mathcal{P}\times\mathcal{T}$ symmetry then it must have $\boldsymbol{\mathcal{B}}_n(\textbf{k}) = 0$ throughout the BZ.

\begin{figure}
\vspace*{-3.2cm}
\setlength{\abovecaptionskip}{-60pt plus 0pt minus 0pt}
\centering
\includegraphics[width=0.66\textwidth, angle=90]{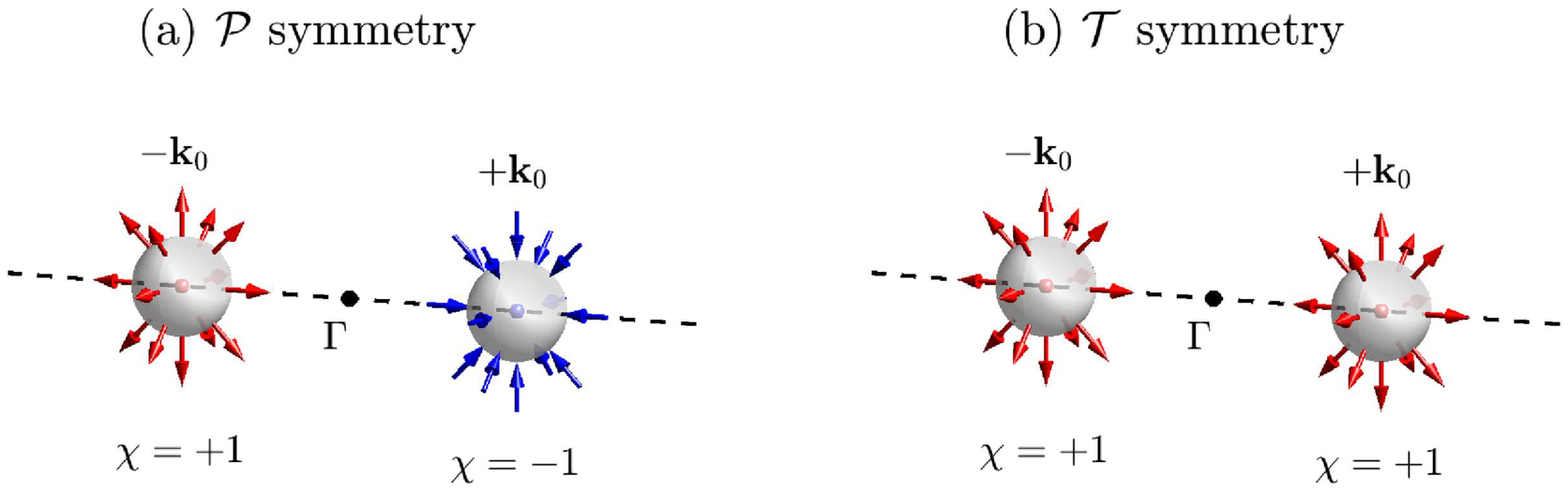}\vspace{0cm}
\caption{(a) Berry curvature field for the lower energy band in the vicinity of a pair of Weyl nodes in a crystal with inversion ($\mathcal{P}$) symmetry but broken time-reversal ($\mathcal{T}$) symmetry. The field satisfies the symmetry relation $\boldsymbol{\mathcal{B}}_n(\textbf{k}) = \boldsymbol{\mathcal{B}}_n(-\textbf{k})$. (b) The same for a crystal with $\mathcal{T}$ symmetry but broken $\mathcal{P}$ symmetry, for which $\boldsymbol{\mathcal{B}}_n(\textbf{k}) = -\boldsymbol{\mathcal{B}}_n(-\textbf{k})$. The grey spheres represent a closed surface over which the surface integral in Equation~(\ref{eq:Chern_thm}) can be calculated in order to determine the Chern number, $C_n$. The sign of $C_n$ for the lower energy band gives the chirality ($\chi$) of the node, as indicated. The point $\textbf{k} = 0$ is marked $\Gamma$.}\label{fig:Chirality}
\end{figure}

The transformation properties of $\boldsymbol{\mathcal{B}}_n(\textbf{k})$ place constraints on the location of Weyl nodes in \textbf{k} space and their chirality.  If a crystal with either inversion or time-reversal symmetry has a Weyl node at $\textbf{k}_0$, then the relation $E(\textbf{k}) = E(-\textbf{k})$ (see Section~\ref{subsec:symmetry-bands}) requires that there must also be a Weyl node at $-\textbf{k}_0$. In the case of inversion symmetry, the condition $\boldsymbol{\mathcal{B}}_n(\textbf{k}) = \boldsymbol{\mathcal{B}}_n(-\textbf{k})$ means that the nodes at $\pm\textbf{k}_0$ will have opposite chirality, whereas for time-reversal symmetry they have the same chirality because $\boldsymbol{\mathcal{B}}_n(\textbf{k}) = -\boldsymbol{\mathcal{B}}_n(-\textbf{k})$. These two situations are illustrated in Figure~\ref{fig:Chirality}. If neither $\mathcal{P}$ nor $\mathcal{T}$ symmetry is present, then a Weyl node at $\textbf{k}_0$ will not in general have a partner at $-\textbf{k}_0$.

It follows from the fermion doubling theorem, Section~\ref{sec:Fermion-doubling-theorem}, that the minimum number of Weyl nodes in a crystal with $\mathcal{P}$ symmetry but broken $\mathcal{T}$ symmetry is two, since in this case the chirality of the $\pm\textbf{k}_0$ nodes is opposite and sums to zero. On the other hand, the minimum number of nodes in a crystal with $\mathcal{T}$ symmetry but broken $\mathcal{P}$ symmetry is four, because the $\pm\textbf{k}_0$ nodes have the same chirality and so there must be at least one other pair of nodes with the opposite chirality to satisfy the theorem. The nodes with positive chirality will in general be at a different energy to the pair with negative chirality.

The properties of Weyl nodes under different combinations of $\mathcal{P}$ and $\mathcal{T}$ symmetry are summarised in Table~\ref{Table1}.

\begin{table}[b!]
\tbl{Effect of space inversion ($\mathcal{P}$) and time-reversal ($\mathcal{T}$) symmetry on some properties of electronic bands and Weyl points (WPs).}
{\begin{tabular}{lcccc} \toprule
& \multicolumn{2}{l}{\hspace{4cm}Type of symmetry present} \\ \cmidrule{2-5}
Property & $\mathcal{P}$ & $\mathcal{T}$ & $\mathcal{P}$ and $\mathcal{T}$ & $\mathcal{P} \times \mathcal{T}$\textsuperscript{a} \\ \midrule
Band dispersion & $E_n(\textbf{k}) = E_n(-\textbf{k})$ &  $E_n(\textbf{k}) = E_n(-\textbf{k})$ &  $E_n(\textbf{k}) = E_n(-\textbf{k})$ &  $E_n(\textbf{k}) \ne E_n(-\textbf{k})$  \\[2pt]
Berry curvature & $\boldsymbol{\mathcal{B}}_n(\textbf{k}) = \boldsymbol{\mathcal{B}}_n(-\textbf{k})$ & \hspace{2.4mm}$\boldsymbol{\mathcal{B}}_n(\textbf{k}) = -\boldsymbol{\mathcal{B}}_n(-\textbf{k})$ & $\boldsymbol{\mathcal{B}}_n(\textbf{k}) = 0$ & $\boldsymbol{\mathcal{B}}_n(\textbf{k}) = 0$  \\[2pt]
Chirality of WPs & \hspace{2.3mm}$\chi(\textbf{k}_0) = -\chi(-\textbf{k}_0)$ & $\chi(\textbf{k}_0) = \chi(-\textbf{k}_0)$ & no WPs & no WPs  \\[2pt]
Min. no. of WPs & 2 & 4 & -- & --  \\ \bottomrule
\end{tabular}}
\tabnote{\textsuperscript{a} Both $\mathcal{P}$ and $\mathcal{T}$ symmetries are broken, but their combination $\mathcal{P} \times \mathcal{T}$ (or $\mathcal{T} \times \mathcal{P}$) is present.}
\label{Table1}
\end{table}

\section{Two-dimensional topological band insulators and edge states}\label{sec:2DTI}

The first accounts of topological effects in electronic insulators on a lattice were based on model two-dimensional (2D) systems \cite{TKNN1982,Haldane1988,KaneMele2005a,Bernevig2006}. In 2D, there is only one closed surface in \textbf{k} space and that is the entire Brillouin zone, which is closed because of the periodicity of the Bloch functions in reciprocal space (see Figure~\ref{fig:BZ}).  Hence, each isolated band (i.e.~one that doesn't touch any other bands) has a Chern number given by
\begin{equation}
C_n = \frac{1}{2\pi}\int_\textrm{BZ} \mathcal{B}_n^{xy}(\textbf{k})\,\textrm{d}^2\textbf{k}.
\label{eq:Chern_2D}
\end{equation}
In 2D, the Berry curvature is a scalar [c.f.~the 3D case, Equation~(\ref{eq:Berry-curvature-curl})],
\begin{align}
\mathcal{B}_n^{xy} & = \frac{\partial A_n^y}{\partial k_x} - \frac{\partial A_n^x}{\partial k_y}.
\label{eq:Berry-curvature_2D}
\end{align}
A band in 2D can have a non-zero Chern number only if the Hamiltonian has broken time-reversal symmetry, because time-reversal symmetry requires  $\mathcal{B}_n^{xy}$ to be an odd function of \textbf{k} (see Section~\ref{subsec:symmetry-Berry}).

\subsection{Quantum anomalous Hall insulator}\label{subsec:QAH}

Two-dimensional electron systems with $C_n \ne 0$ are called \emph{Chern insulators}, or alternatively \emph{quantum anomalous Hall (QAH) insulators}. This class of material was introduced by Haldane in 1988 \cite{Haldane1988} in order to show how the quantum Hall effect can arise on a lattice in the absence of an applied magnetic field. The key ingredients of Haldane's model are (i) a band inversion, and (ii) a type of magnetic ordering that is lattice periodic but has no net magnetic flux through the surface and creates an insulating gap. The QAH effect was first observed in thin films of (Bi,Sb)$_2$Te$_3$ doped with Cr to introduce ferromagnetism \cite{Chang2013}.

QAH insulators have a quantised Hall conductivity given by $\sigma_{yx} = Ce^2/h$, where $C = \sum_{n \le n_\textrm{F}} C_n$ is the total Chern number for the occupied bands. They have an edge band of states that crosses the gap between the bulk valence and conduction bands at the boundary where the gap closes and the topology changes, as described in Section~\ref{sec:topological-insulators}. The edge states are topologically protected, which means that they are not eliminated by small perturbations, such as lattice distortions or non-magnetic impurities. Their existence is guaranteed as long as there is a change in topology on crossing the boundary, as illustrated in Figure~\ref{fig:TI}. Moreover, because of the broken time-reversal symmetry, the currents associated with the edge mode flow around the boundary in one direction only, that direction being given by the gradient (group velocity) of the 1D surface band of states. For this reason, the edge modes are termed \emph{chiral}. Note, however, that the edge modes are not, in general, fully spin-polarised.

\begin{figure}
\vspace*{-2cm}
\setlength{\abovecaptionskip}{-20pt plus 0pt minus 0pt}
\centering
\includegraphics[width=0.6\textwidth, angle=90]{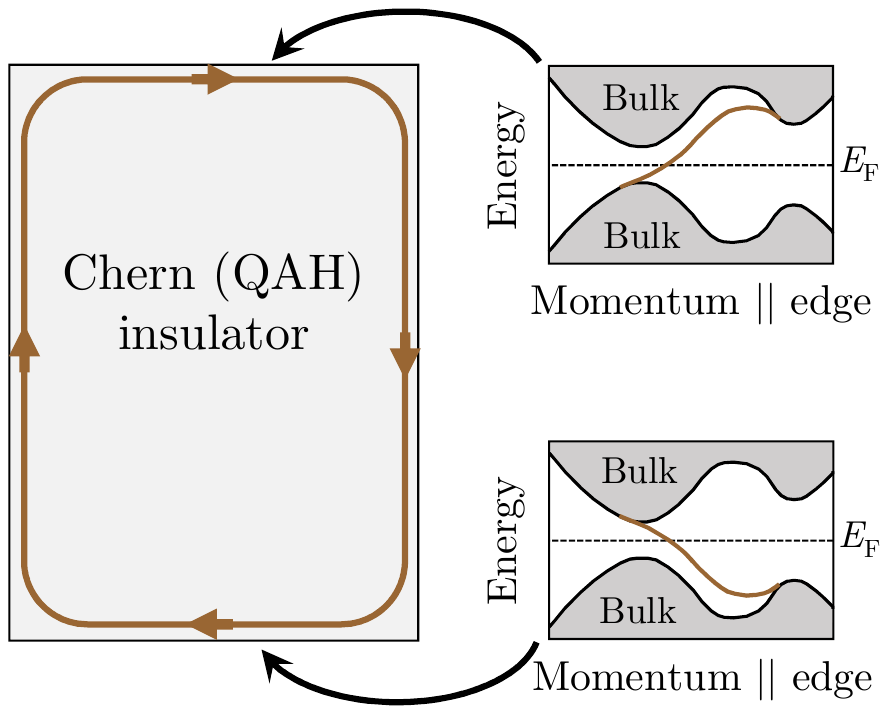}\vspace{0cm}
\caption{Left: chiral edge state on a Chern (QAH) insulator. Right: energy spectrum as a function of momentum along the top and bottom edges. The shaded regions are the projections of the bulk valence and conduction bands. A single edge band connects the bulk valence and conduction bands with a positive (negative) Fermi velocity on the top (bottom) edges. } \label{fig:QAH}
\end{figure}

Figure~\ref{fig:QAH} is a schematic representation of a simple band of states on the edge of a QAH insulator. The band dispersion on the bottom edge is reversed relative to the top edge, corresponding to the opposite velocity. The displayed band crosses the Fermi energy $E_\textrm{F}$ only once, but this needn't be the case.  An edge band could in principle cross $E_\textrm{F}$ multiple times, generating left- and right-moving states at $E_\textrm{F}$. In general, the number of times an edge band crosses the Fermi level is governed by the \emph{bulk--boundary correspondence}, which can be expressed as
\begin{align}
N_\textrm{R} - N_\textrm{L} = \Delta C_n,
\label{eq:bulkboundary}
\end{align}
where $N_\textrm{R}$ and $N_\textrm{L}$ are the number of right- and left-moving modes, and $\Delta C_n$ is the change in Chern number across the boundary. The band shown in Figure~\ref{fig:QAH}, therefore, corresponds to a change in $C_n$ from 1 to 0.

\subsection{Quantum spin Hall insulator}\label{subsec:QSH}

\begin{figure}
\vspace*{-2cm}
\setlength{\abovecaptionskip}{-20pt plus 0pt minus 0pt}
\centering
\includegraphics[width=0.6\textwidth, angle=90]{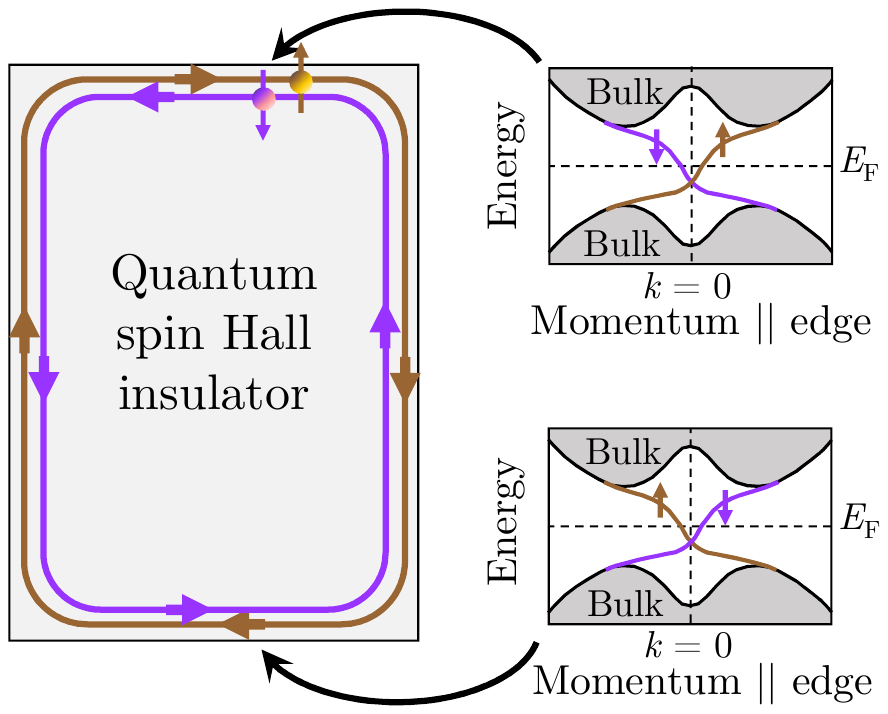}\vspace{0cm}
\caption{Left: helical (spin-polarised) edge state on a Quantum spin Hall (QSH) insulator.  Right: energy spectrum as a function of momentum along the top and bottom edges. The shaded regions are the projections of the bulk valence and conduction bands. A pair of edge band connects the bulk valence and conduction bands with opposite Fermi velocities. The band crossing at the TRIM $(k=0)$ is protected by time-reversal symmetry.} \label{fig:QSH}
\end{figure}

In the QAH insulator, the topological effects require magnetic order. In 2005, Kane and Mele introduced an important model which showed that topological effects can be present even without magnetic order \cite{KaneMele2005a,KaneMele2005b}. Their model is similar to Haldane's but includes spin-orbit coupling (SOC) in place of magnetic order.  A key difference is that the SOC interaction is time-reversal invariant, whereas magnetic order is not.  The SOC ensures that there is a gap at points where the bulk bands cross. Remarkably, when the parameters of the model are tuned to produce a band inversion, pairs of singly degenerate spin-polarised, or \emph{helical}, edge states appear which cross the gap between the bulk valence and conduction bands, see Figure~\ref{fig:QSH}. By \emph{helical} we mean that the up and down spin states travel in opposite directions around the boundary. The associated currents are unaffected by non-magnetic perturbations because there is no conduction channel for electrons whose direction of motion is reversed but whose spin is unchanged.

The helical surface bands cross at the TRIM of the surface BZ, and the degeneracy is protected by time-reversal symmetry (Section~\ref{subsec:symmetry-bands}). This ensures a metallic surface. If $\mathcal{T}$-invariance is broken, e.g.~by magnetic order, then gaps open at the TRIM and the surface will become insulating if the gap is large enough to contain $E_\textrm{F}$.

The type of system described by the Kane--Mele model is termed a \emph{quantum spin Hall (QSH) insulator}, or simply a \emph{2D topological insulator}. From a topological point of view, the QSH insulator is interesting because the presence of time-reversal symmetry implies $C_n = 0$ (Section~\ref{subsec:symmetry-Berry}), and so there must be another topological invariant that distinguishes a QSH insulator from a normal band insulator.

To see how this comes about, let us label the up- and down-spin bands $\uparrow$ and $\downarrow$. Because time-reversal transforms $\uparrow$ to $\downarrow$, the states in these bands are not time-reversal invariant and so have separate Chern numbers $(C_{n\uparrow}, C_{n\downarrow})$ which do not have to be zero providing $C_n = C_{n\uparrow} + C_{n\downarrow} = 0$. It is found that when $C_{n\uparrow}$ and $C_{n\downarrow}$ are both odd integers the system is a QSH insulator, and when they are both even integers it is a normal insulator.  This type of topology is characterised by a $\mathbb{Z}_2$ topological invariant \cite{KaneMele2005b}, where $\mathbb{Z}_2$ represents the group of integers $(0,1)$ under addition modulo 2. It can be shown that the $\mathbb{Z}_2$ index (0 or 1) measures the number of times (mod 2) an edge band dispersion crosses the Fermi level in half of the 1D surface BZ, i.e.~$0 \le k \le \pi/a$, so it provides a simple test to distinguish an insulator which is topologically trivial ($\mathbb{Z}_2$-even) from one that is topologically non-trivial ($\mathbb{Z}_2$-odd). The edge bands shown in Figure~\ref{fig:QSH} cross $E_\textrm{F}$ once in the half BZ, corresponding to a topologically non-trivial insulator.

\section{3D topological insulators}\label{sec:Insulators}

Three-dimensional (3D) topological insulators (TIs) are insulators in the bulk but have conducting states on some or all of their surfaces (Section~\ref{sec:topological-insulators}). They are closely related to the 2D insulators described in Section~\ref{sec:2DTI}, in which  time-reversal symmetry is either present (QSH insulator) or broken (QAH/Chern insulator).

\subsection{Time-reversal symmetric TIs}\label{sec:3DTIs}

\begin{figure}
\vspace*{-4cm}
\hspace*{-3pt}
\setlength{\abovecaptionskip}{-80pt plus 0pt minus 0pt}
\centering
\includegraphics[width=0.8\textwidth, angle=90]{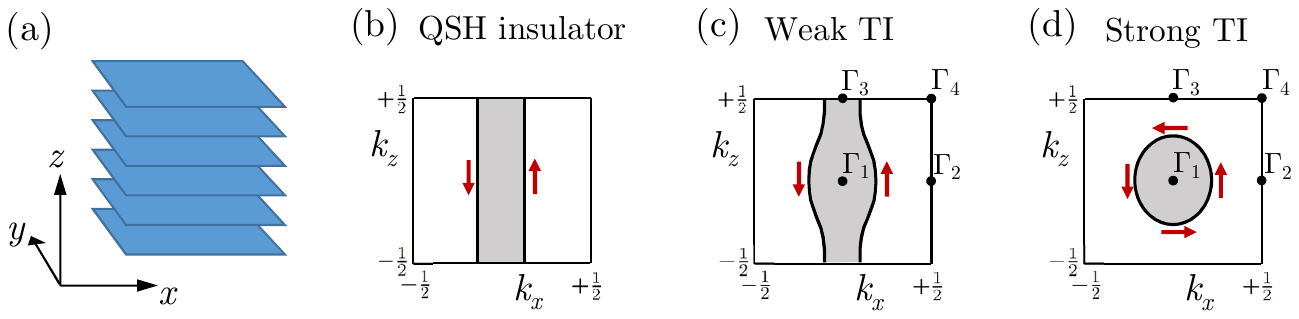}\vspace{0cm}
\caption{(a) Stack of QSH insulators. (b)--(d) Surface Brillouin zone and Fermi energy contour for 3D TIs with increasing inter-layer couplings. The shaded regions indicate filled states, the red arrows represent the spin direction, and $\Gamma_{1,2,3,4}$ are the TRIM. (b) zero coupling; (c) weak TI --- the Fermi contour encloses two TRIM ($\nu_0 = 0$); (d) strong TI --- the Fermi contour encloses one TRIM ($\nu_0 = 1$).} \label{fig:TI3D}
\end{figure}

Consider a 3D material built out of 2D QSH insulating layers stacked on top of one another in the $z$ direction like a deck of cards, as shown in Figure~\ref{fig:TI3D}(a). An isolated 2D QSH layer has helical edge states as described in Section~\ref{subsec:QSH}. In the simplest case (Figure~\ref{fig:QSH}), the bulk band gap is traversed by a single pair of Kramers-degenerate edge states, $E(-k) = E(k)$, which cross the Fermi level ($E_\textrm{F}$) once in the half BZ $k > 0$.  If the coupling between adjacent QSH layers in the stack is zero, then there is no dispersion in the inter-layer direction $k_z$, and the Fermi energy contour that separates the filled and empty surface states is a pair of straight lines running parallel to $k_z$, as shown in Figure~\ref{fig:TI3D}(b).

Now suppose that there is a small but non-zero inter-layer coupling. The surface bands may develop a weak dispersion along $k_z$, giving a slightly curved Fermi contour, Figure~\ref{fig:TI3D}(c), which remains open, i.e.~the contour connects to itself on opposite boundaries of the BZ. Finally, with a stronger inter-layer coupling we might obtain a closed Fermi contour like that sketched in Figure~\ref{fig:TI3D}(d).

Figures~\ref{fig:TI3D}(c) and (d) are illustrative of what are known as \emph{weak} and \emph{strong} TIs. The distinction can be understood if we consider how the bands connect the time-reversal invariant momenta (TRIM) in the BZ. Recall that there are four TRIM in the 2D BZ, labelled $\Gamma_1$ to $\Gamma_4$ in Figure~\ref{fig:TI3D}, and that surface states, if they exist, must be Kramers degenerate at these points. Elsewhere, the degeneracy is lifted by spin--orbit coupling, so that the TRIM are the nodes of a 2D linear band crossing like the Dirac dispersion in graphene (see Section~\ref{subsec:graphene}).

In Figure~\ref{fig:TI3D}(c), a surface band crosses $E_\textrm{F}$ once along the path from $\Gamma_1$ to $\Gamma_2$ ($k_x$ direction) but does not cross $E_\textrm{F}$ between $\Gamma_1$ and $\Gamma_3$ ($k_z$ direction). In Figure~\ref{fig:TI3D}(d), a band crosses $E_\textrm{F}$ once along both paths. Therefore, situations (c) and (d) are not smoothly connected and represent distinct topological phases. In particular, in (c) the surface perpendicular to $k_z$ is gapped and insulating, like the original QSH insulator, whereas in (d) it is metallic. The surfaces perpendicular to $k_x$ and $k_y$ are metallic in both cases.

More precisely, a weak TI is one in which the Fermi contour encloses an even number of TRIM, whereas in a strong TI it encloses an odd number. This classification is represented by a $\mathbb{Z}_2$ topological index $\nu_0$, such that a weak TI has $\nu_0 = 0$ (even integer, modulo 2), and a strong TI has $\nu_0 = 1$ (odd integer, modulo 2). The value of $\nu_0$ depends on symmetry properties of the Bloch functions at all eight TRIM of the bulk BZ. More generally, the Fermi contour may intersect the line joining any pair of TRIM an odd or an even number of times, and in order to describe all cases three more $\mathbb{Z}_2$ indices, $\nu_1$, $\nu_2$ and $\nu_3$ are required \cite{MooreBalents2007,FuKaneMele2007,Roy2009}.

A magnetic field applied perpendicular to the conducting surface of a 3D TI will generate Landau levels, as for a conventional 2D metal, resulting in quantum oscillations and an associated quantum Hall effect. The Hall conductivity $\sigma_{xy}$ is quantised in steps of $e^2/h$, as usual, but the first step is at $e^2/2h$. This half-quantisation is a result of the $\pi$ Berry phase for adiabatic transport around a Dirac point in 2D (Section~\ref{subsec:2-level}). It is important to appreciate, however, that the surface states of a 3D TI are fundamentally different from a topologically trivial 2D metal. In the latter, there are up and down spins at every point on the Fermi contour, whereas in a 3D TI, time-reversal symmetry requires that $+\textbf{k}$ and $-\textbf{k}$ states have opposite spin, so the spin must rotate with $\textbf{k}$ around the Fermi contour, Figure~\ref{fig:TI3D}(c) and (d). This helical spin texture has been observed in ARPES experiments \cite{Hsieh2009}.

\subsection{3D magnetic TIs}\label{sec:3DChern}

\begin{figure}
\vspace*{-4cm}
\setlength{\abovecaptionskip}{-80pt plus 0pt minus 0pt}
\centering
\includegraphics[width=0.7\textwidth, angle=0]{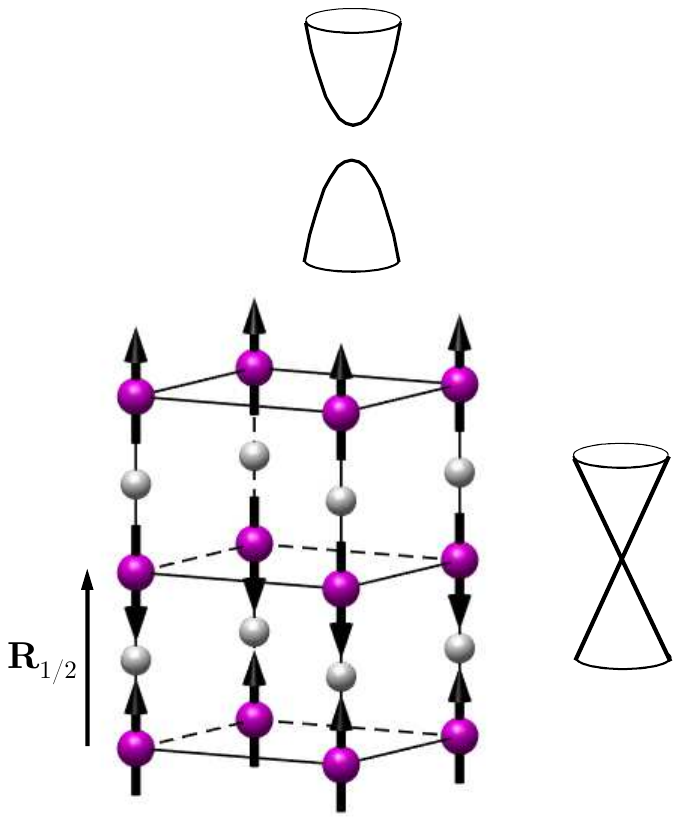}\vspace{0cm}
\caption{An antiferromagnetic spin structure which has $S = \mathcal{T} \times \textbf{R}_{1/2}$ symmetry, where $\textbf{R}_{1/2}$ is a half-lattice translation in the vertical direction which connects sites with opposite spin. A strong topological insulator with this AFM structure is an example of an axion insulator. As indicated, the Dirac cones on the top and bottom surfaces of the crystal are gapped, whereas those on the sides are gapless.} \label{fig:Axion}
\end{figure}

Much of the research on 3D TIs has focussed on time-reversal-invariant systems \cite{HasanKane2010,Ando2013}. Magnetic TIs, which combine non-trivial band topology and magnetic order, are expected to exhibit some similar and some distinct features compared with their non-magnetic counterparts \cite{Tokura2019,Nenno2020,Sekine2021,BernevigFelserBeidenkopf2022,Wang2023}. Magnetic TIs offer the added possibility of magnetic control of their topological states. To work with such materials requires a detailed knowledge of their magnetic structures, which can be obtained from neutron diffraction and x-ray resonant scattering measurements (see, for example, Refs.~\cite{Soh2021,Rahn2018}).

A simple way to create a 3D magnetic TI is to introduce ferromagnetism into a 3D $\mathcal{T}$-invariant TI, either by doping or a ferromagnetic coating layer. The magnetisation \textbf{M} couples to the surface states via an exchange interaction, and opens a gap in the Dirac surface spectrum of the host TI on the boundary surfaces perpendicular to \textbf{M}. The sign of the Chern number $C$ $( =\pm 1)$ depends on the direction of \textbf{M}. On entering the material across the side surfaces parallel to \textbf{M}, $C$ jumps from 0 to $\pm 1$, so a dissipationless chiral edge mode appears in accordance with the bulk--boundary correspondence, just as for the 2D QAH insulator. If $E_\textrm{F}$ lies in the gap then $\sigma_{xy}$ is half-quantised (N.B.: in a measurement that probes both top and bottom samples surfaces, the observed $\sigma_{xy}$ will be $\pm e^2/h$, i.e.~$\pm e^2/2h$ from each surface).

One problem with doping is that it leads to inhomogeneous electronic properties. Recent research, therefore, has focussed on intrinsic (stoichiometric and well-ordered) magnetic compounds.  Pioneering work reported in 2010 \cite{Mong2010} showed that a strong TI with broken $\mathcal{T}$ symmetry due to a type of antiferromagnetic (AFM) order is topological in a similar sense to $\mathcal{T}$-invariant TIs, but has some surfaces which are gapped and some which are not gapped. The AFM order considered in Ref.~\cite{Mong2010} has twice the periodicity of the crystal lattice but preserves the combined symmetry $S = \mathcal{T} \times \textbf{R}_{1/2}$, where $\textbf{R}_{1/2}$ is a half-lattice translation which connects sites with opposite spin. A simple example is shown in Figure~\ref{fig:Axion}.  Such an AFM crystal has two types of surfaces, those in which all the spins are aligned ferromagnetically (the top and bottom in Figure~\ref{fig:Axion}), and those in which there is AFM order within the surface (the sides). The latter has the $S$ symmetry, which preserves the Kramers degeneracy of the surface states and preserves an odd number of gapless Dirac cones, as for a strong TI. Ferromagnetic surfaces, on the other hand, break $S$ symmetry because the $\textbf{R}_{1/2}$ translation is terminated at the surface, and so these surfaces are gapped and exhibit a half-integer $\sigma_{xy}$.

Magnetic TI materials whose bulk symmetries include $\mathcal{T}$ combined with another crystalline symmetry to give topological properties similar to those just described are termed \emph{axion insulators} \cite{Nenno2020,Sekine2021}. The name comes from a particular type of quantised magnetoelectric coupling associated with the electrodynamics of TIs. The coupling, which adds a term $\mathcal{L}_\theta \propto \theta \textbf{E}\cdot\textbf{B}$ to the electromagnetic Lagrangian of the material, where $\theta = 0$ (trivial insulator) or $\pm \pi$ (topological insulator), is analogous to the axion field introduced in high-energy physics to explain the absence of charge--parity violation in the strong interaction between quarks \cite{Wilczek1987}. The first intrinsic AFM axion insulator material to be identified was MnBi$_2$Te$_4$ \cite{Otrokov2019}.

\section{Nodal semimetals}\label{sec:Dirac-Weyl}

\subsection{Weyl semimetals}\label{sec:WSM}

\begin{figure}
\vspace*{-1cm}
\setlength{\abovecaptionskip}{0pt plus 0pt minus 0pt}
\centering
\includegraphics[width=0.3\textwidth, angle=0]{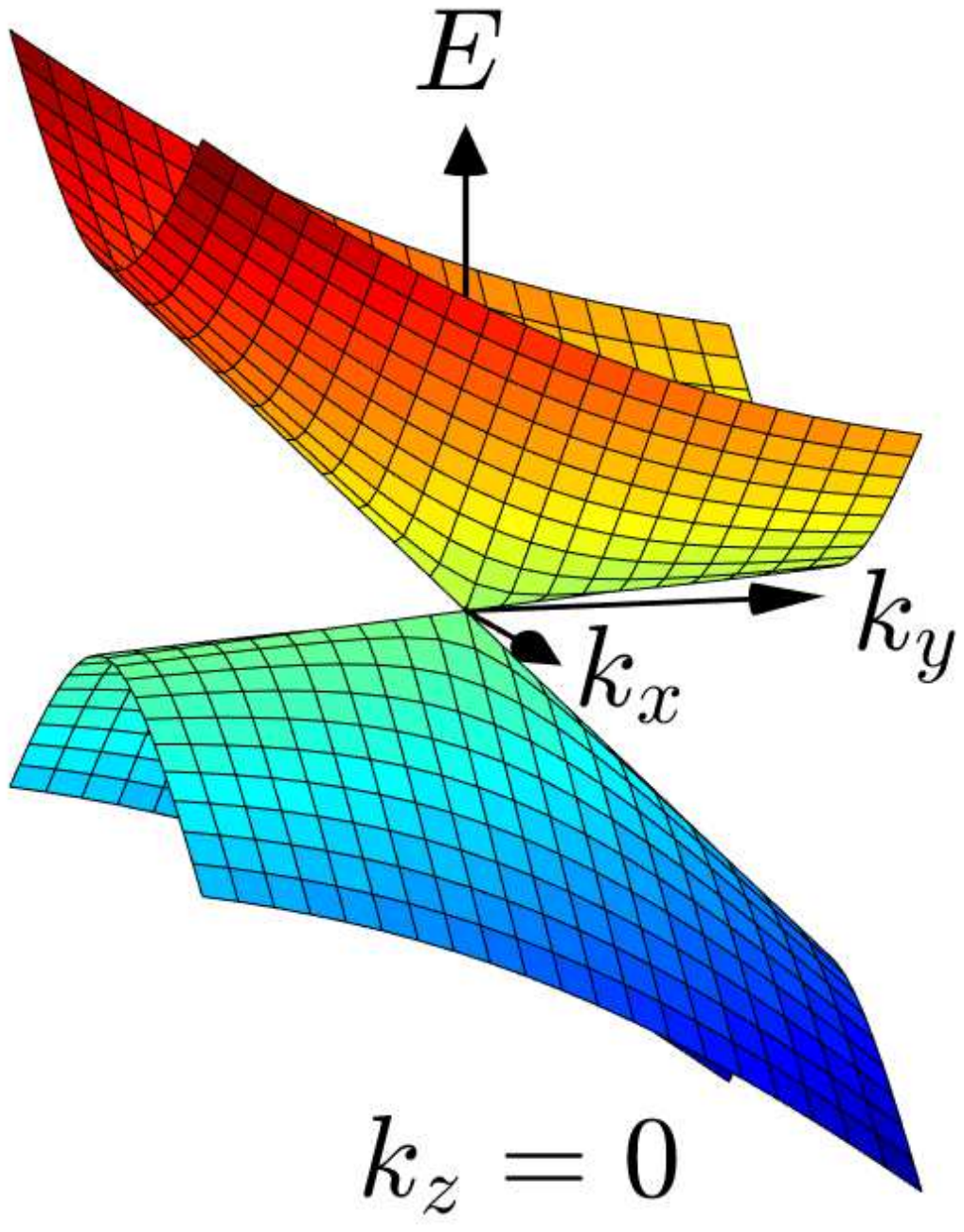}\vspace{0cm}
\caption{Dispersion surface in the vicinity of an anisotropic and tilted band crossing at $\textbf{k} = 0$.} \label{fig:Tilted}
\end{figure}

Recall that a Weyl point (or node) is a touching point between two singly degenerate bands in a 3D crystal (Section~\ref{subsec:2-level}). In Section~\ref{subsec:symmetry-Berry}, we showed that the existence of Weyl nodes requires either broken $\mathcal{P}$ symmetry, or broken $\mathcal{T}$ symmetry, or both.  In Section~\ref{subsec:2-level}, we analysed a simple model Hamiltonian that describes a Weyl node with a linear isotropic band dispersion. We assigned a chirality $\chi = \pm 1$ to the Weyl point depending on whether the lower band is a source or sink of Berry curvature, and the results in Section~\ref{sec:Chern-theorem} show that a Weyl point is topological because the Chern number $C$ calculated for a closed surface surrounding the node is non-zero. The only way to eliminate Weyl points is for pairs with opposite chirality to annihilate one another in order to preserve zero net chirality, as required by the Nielsen-Ninomiya theorem, Equation~(\ref{Nielsen-Ninomiya-thm}). Annihilation requires the pair to move to the same \textbf{k}, and so Weyl points are stable against small changes in the Hamiltonian as long as they remain separated in \textbf{k} space.

In 3D materials, the dispersion around a touching point between two bands is often anisotropic and tilted, as illustrated in Figure~\ref{fig:Tilted}. We can describe such cases with a more general two-band model whose effective Hamiltonian can be represented by the two-by-two Hermitian matrix
\begin{align}
{\mathcal H}(\textbf{k}) & = \left(\begin{array}{c c} f_0 + f_3 & f_1 - if_2\\[2pt] f_1 +if_2 & f_0 - f_3 \end{array} \right) \label{eq:Weyl-H-gen-1} \\[10pt] & = f_0\sigma_0 +  f_1\sigma_1 +f_2\sigma_2 + f_3\sigma_3,
\label{eq:Weyl-H-gen-2}
\end{align}
where $f_0$ to $f_3$ are real functions of \textbf{k}. On the second line, $\sigma_0$ is the $2\times 2$ identity matrix, and $\sigma_1$, $\sigma_2$ and $\sigma_3$ are the Pauli matrices. The eigenvalues of (\ref{eq:Weyl-H-gen-1}) are $E_{\pm} = f_0 \pm \sqrt{f_1^2+f_2^2+f_3^2}$, so the band degeneracy occurs when $f_1 = f_2 = f_3 = 0$.

If, for example, we put $(f_1, f_2, f_3) = \hbar (v_x k_x, v_y k_y, v_z k_z)$, then we obtain an anisotropic linear dispersion, with speeds $(v_x, v_y, v_z)$ along a set of orthogonal axes $(k_x, k_y, k_z)$. If, in addition, we put $f_0 = \hbar \textbf{v}_0\cdot \textbf{k}$, then we obtain a tilt in the direction of $\textbf{v}_0$.   The linear isotropic Hamiltonian, Equation (\ref{eq:H(k)-1}), is recovered when $(f_0, f_1, f_2, f_3) = \hbar v (0, k_x, k_y, k_z)$. An anisotropic and tilted linear dispersion can always be brought into diagonal isotropic form by an orthogonal transformation together with a scaling of the $k_i$ axes. Hence, the topology is the same as the isotropic case, and the Berry curvature derives from monopoles of strength $\pm\frac{1}{2}$ in the diagonal/scaled basis.

Weyl points are not restricted to any particular energy, but if they are located close to the Fermi energy then they can influence the excitations which are responsible for the low-energy properties of the material. Consider once again the case of a linear isotropic band crossing at $\textbf{k} = \textbf{k}_0$ with the zero of energy at the Weyl node, and put $\textbf{q} = \textbf{k} - \textbf{k}_0$. The second form of the effective Hamiltonian, given in Equation~(\ref{eq:Weyl-H-gen-2}), may then be written
\begin{align}
{\mathcal H}(\textbf{q}) = \pm\hbar v \textbf{q}\cdot \mbox{\boldmath $ \sigma$},
\label{eq:Weyl-H-gen-3}
\end{align}
where $\mbox{\boldmath $ \sigma$} = (\sigma_1, \sigma_2, \sigma_3)$, and the $\pm$ sign describes the two different chiralities of Weyl point.

Equation~(\ref{eq:Weyl-H-gen-3}) takes exactly the same form, up to the replacement of $v$ by the speed of light $c$, as the Hamiltonian proposed by Hermann Weyl in 1929 for massless relativistic fermions \cite{Weyl1929a,Weyl1929b}. Therefore, in a crystal whose electronic states are described by (\ref{eq:Weyl-H-gen-3}) the low energy quasiparticle excitations mimic Weyl fermions, except they travel at speeds $v \ll c$. We call them \emph{emergent Weyl fermions}. It is important to note that in the case of elementary particles, $\mbox{\boldmath $ \sigma$}$ represents the spin of the particle, whereas for the emergent quasiparticles $\mbox{\boldmath $ \sigma$}$  is a \emph{pseudospin} which acts in the two-band sub-space of Bloch functions and is not necessarily a pure spin degree of freedom.  Equation~(\ref{eq:Weyl-H-gen-3}) shows that the pseudospin is coupled to the reduced wavevector \textbf{q}, a property known as \emph{helicity}, and the sign determines whether the pseudospin is parallel or antiparallel to \textbf{q}.

If the Weyl node is located right at the Fermi energy $E_\textrm{F}$, then we have what is known as a \emph{Weyl semimetal} (WSM). A semimetal is characterised by a vanishing electronic density of states at $E=E_\textrm{F}$ but non-zero density of states for $E>E_\textrm{F}$ and $E<E_\textrm{F}$.  If the node is just above or just below $E_\textrm{F}$ then the material is termed a \emph{Weyl metal}. A Weyl metal retains many of the characteristics of a WSM providing $E_\textrm{F}$ is located on the linear part of the dispersion, and providing there are no trivial bands at $E_\textrm{F}$ to mask the response of the Weyl fermions.

\subsection{Dirac semimetals}\label{sec:DSM}

A Dirac point forms at the crossing of two doubly degenerate bands, rather than two singly degenerate bands as in the case of a Weyl point. A system in which the Dirac node occurs at the Fermi level is termed a \emph{Dirac semimetal} (DSM). The conical band dispersion of an ideal DSM is the same as that of massless (relativistic) fermions described by the Dirac equation, hence the name.

Dirac points require the combined $\mathcal{P} \times \mathcal{T}$ symmetry in order to enforce double degeneracy throughout the BZ (Section~\ref{subsec:symmetry-bands}), so $\mathcal{P}$ and $\mathcal{T}$ are either both present or both absent. The latter case makes it possible for Dirac points to exist in magnetic materials.

As explained in Section~\ref{subsec:symmetry-Berry},  $\mathcal{P} \times \mathcal{T}$ symmetry means that the Berry curvature $\boldsymbol{\mathcal{B}}_n(\textbf{k})$ is zero throughout the BZ (except at the degeneracy point itself, where $\boldsymbol{\mathcal{B}}_n(\textbf{k})$ is undefined), and so the Chern number of any closed surface surrounding a Dirac point is zero. Hence, a Dirac point can be thought of as the merger of two Weyl nodes of opposite chirality. Nevertheless, Dirac points are topological because they involve band inversion, and there are other topological invariants which characterise them.

When a pair of doubly degenerate bands cross the states can mix, forming a gap in the spectrum at the node. In general, therefore, Dirac points do not form where there is an accidental band crossing at an arbitrary \textbf{k}. Additional symmetries are required to prevent gap formation, for example rotational symmetry axes or high-symmetry points in the BZ, especially TRIM. As a result, Dirac nodes, unlike Weyl nodes, are sensitive to perturbations which remove the symmetry protection of the degeneracy.

\subsection{Graphene}\label{subsec:graphene}

\begin{figure}
\vspace*{-1cm}
\setlength{\abovecaptionskip}{-10pt plus 0pt minus 0pt}
\centering
\includegraphics[width=0.5\textwidth, angle=90]{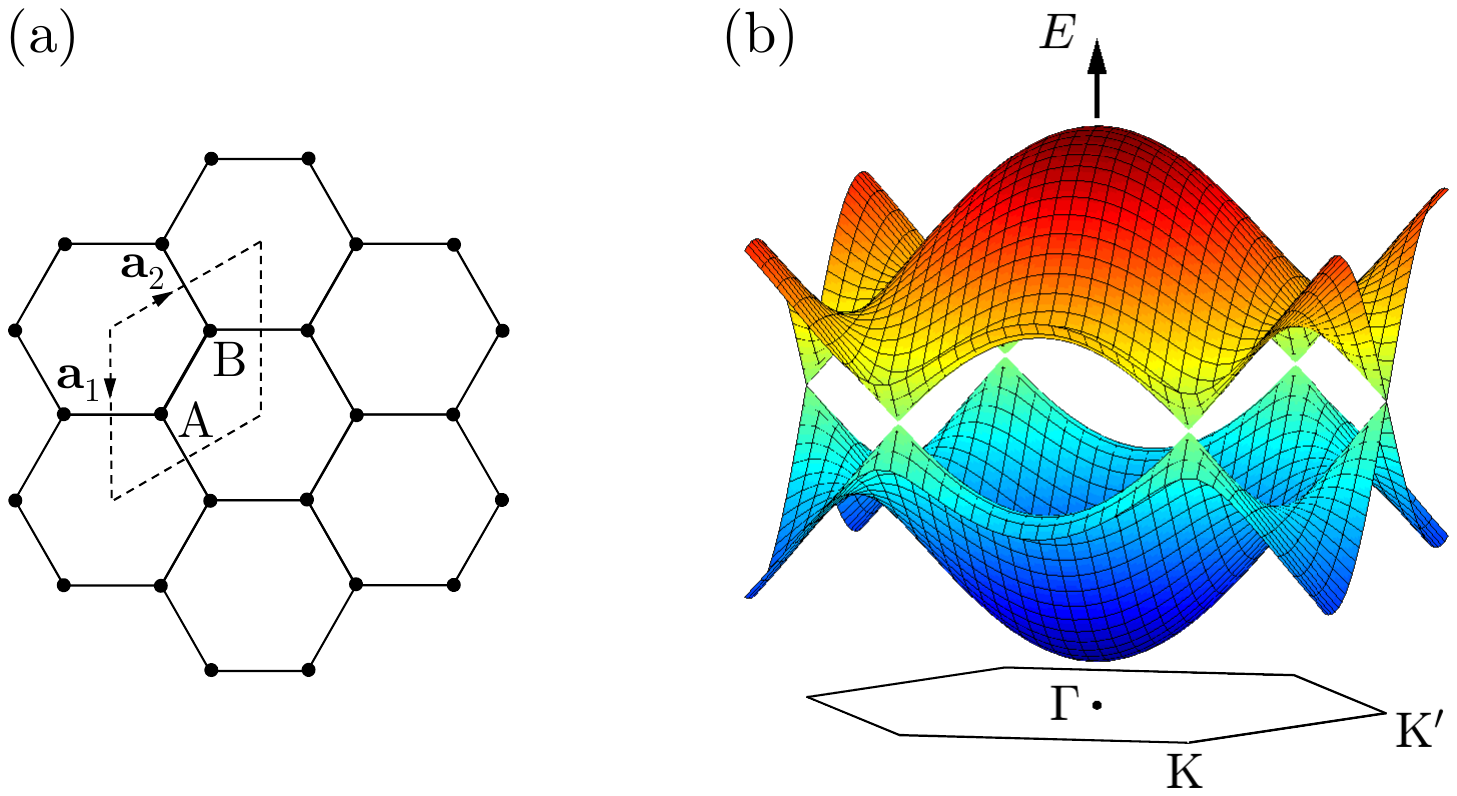}\vspace{0cm}
\caption{(a) Crystal structure of graphene, showing the primitive lattice vectors $\textbf{a}_1$ and $\textbf{a}_2$. A and B label the two carbon atoms in the primitive unit cell. (b) Conduction and valence bands of graphene assuming a two-band tight-binding model with nearest-neighbour coupling $t$.  The BZ is shown below. The band dispersion is given by $E(\textbf{k}) = E_\textrm{F} \pm  t\{1 + 4\cos(3k_xa/2)\cos(\sqrt{3}k_ya/2) + 4\cos^2(\sqrt{3}k_ya/2)\}^{\frac{1}{2}}$, with the Fermi level $E_\textrm{F}$ at the nodes. } \label{fig:graphene}
\end{figure}

To a very good approximation, graphene realises emergent Dirac fermions in 2D, see Figure~\ref{fig:graphene}. The honeycomb arrangement of carbon atoms has $\mathcal{P}$ and $\mathcal{T}$ symmetry, which enforces Kramers degeneracy of the bands. The valence and conduction bands touch at the K and K$'$ points on the BZ boundary, and near to these points the electronic dispersion is linear and isotropic, with a Fermi velocity $v_\textrm{F} \simeq 10^6$\,ms$^{-1}$. The touching points are not TRIM ($\textbf{K}$ and $-\textbf{K}$ are not related by a reciprocal lattice vector), but three-fold rotational symmetry around K and K$'$ fixes the nodes at these points in the BZ. There is in reality a tiny gap ($\sim 10$\,$\mu$eV) due to spin--orbit coupling which turns graphene formally into a quantum spin Hall insulator \cite{KaneMele2005a}, but for most practical purposes the gap can be neglected, and indeed graphene exhibits many of the phenomena predicted by quantum electrodynamics for massless fermions \cite{CastroNeto2009}.

Because of $\mathcal{P}$ and $\mathcal{T}$ symmetry, the Berry curvature $\mathcal{B}^{xy}(\textbf{k})$ of pristine graphene is zero everywhere (except at the band touching points, where it is undefined). Nevertheless, according to Equation~(\ref{eq:Berry-phase-linear}), the Berry phase for one circuit around a K or K$'$ point is $\pm\pi$, and this has been verified experimentally \cite{Novoselov2005}. Therefore, the Bloch states of graphene have non-trivial local geometrical features, but the global topology is trivial.  If $\mathcal{P}$ symmetry is broken with a staggered sublattice potential, making the two carbon sites in the unit cell inequivalent, then a gap opens at the K and K$'$ points and a non-zero $\mathcal{B}^{xy}(\textbf{k})$ emerges, though the Chern number remains zero because as long as $\mathcal{T}$ symmetry is retained $\mathcal{B}^{xy}(\textbf{k})$ is an odd function of \textbf{k}.

\subsection{Surface states and surface Fermi arcs}

\begin{figure}
\vspace*{-1cm}
\setlength{\abovecaptionskip}{-10pt plus 0pt minus 0pt}
\centering
\includegraphics[width=0.5\textwidth, angle=90]{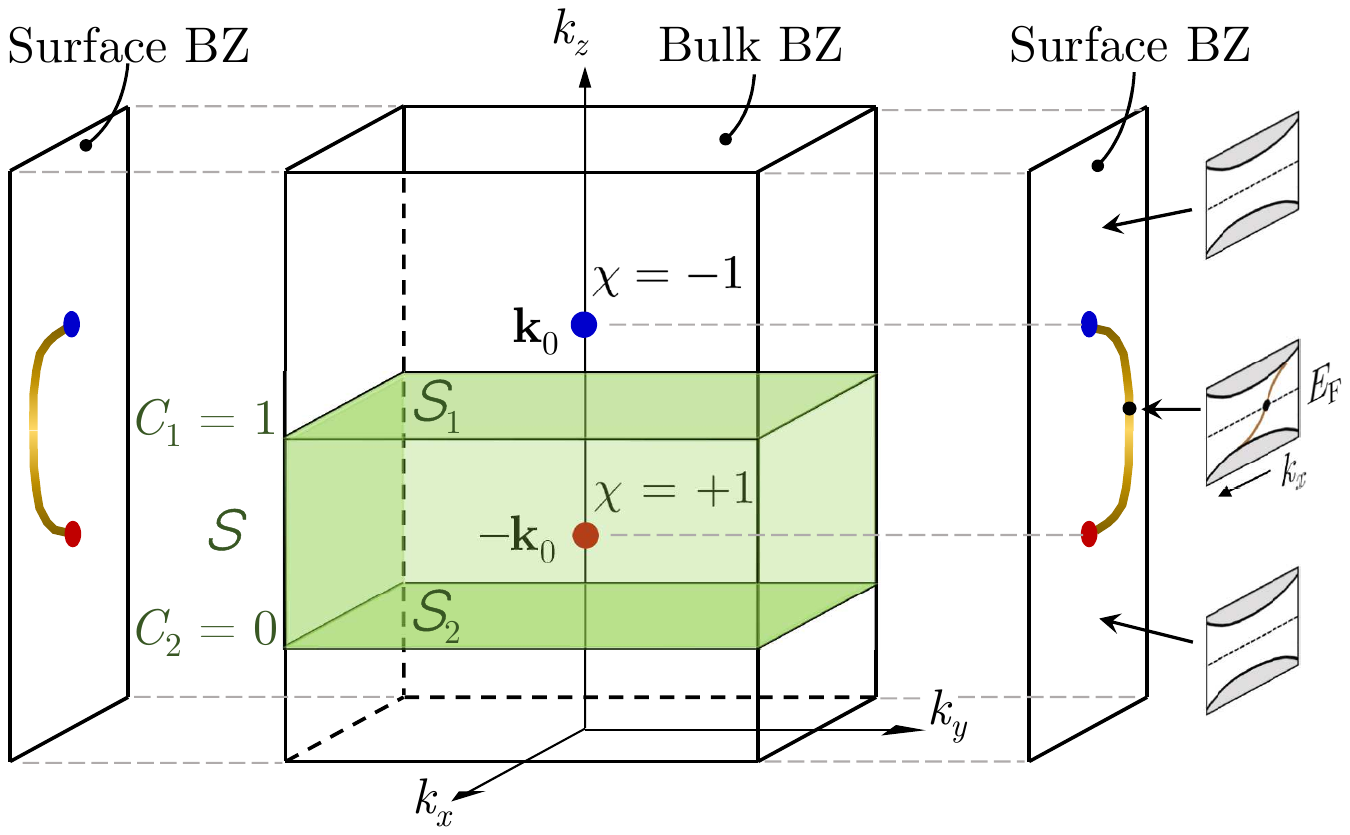}\vspace{0cm}
\caption{Surface Fermi arcs on a Weyl semimetal. Two Weyl points with opposite chirality $\chi = \pm1$ (red and blue circles) are located at wavevectors $\mp\textbf{k}_0$, respectively, in the bulk BZ. The closed surface $\mathcal{S}$ is defined by the pale green planes, and includes the two slice planes $\mathcal{S}_1$ and $\mathcal{S}_2$ together with the side boundaries of the BZ. The Chern numbers of $\mathcal{S}_{1,2}$ are $C_{1,2}$. Fermi arcs are shown on the 2D BZs for the right and left surfaces. On the right side, the energy spectrum is plotted as a function of $k_x$ at three different $k_z$. When $-k_0<k_z<k_0$, a single surface band connects the bulk valence and conduction bands (shaded). The Fermi arc is formed from the locus of $k_x$ points where the surface band crosses $E_\textrm{F}$.} \label{fig:FermiArc}
\end{figure}

We have seen that changes in bulk topology are accompanied by surface or interface states, a property known as the \emph{bulk--boundary correspondence}. The surface states associated with Weyl and Dirac points take the form of Fermi arcs in the 2D surface BZ, and have been extensively studied by angle-resolved photoemission spectroscopy \cite{Lv2021}.

In order to understand the formation of Fermi arcs in more detail, we shall consider the particular case of a $\mathcal{T}$-broken Weyl semimetal with a single pair of Weyl nodes located at wavevectors $-\textbf{k}_0$ and $+\textbf{k}_0$ with chiralities $\chi = +1$ and $-1$, respectively, as shown in Figure~\ref{fig:FermiArc}. We shall assume that the two bands that touch at the Weyl points are separated by an energy gap everywhere else in the BZ, and that the Fermi energy lies at the Weyl nodes. For simplicity, we also assume that the crystal axes are orthogonal, and take $\textbf{k}_0 = (0, 0, k_0)$. We showed in Section~\ref{subsec:2-level} that there is a Chern number associated with a closed surface $\mathcal{S}$ surrounding a Weyl node, and that the Chern number of the lower band is equal to $\chi$.

Now consider $\mathcal{S}$ as shown in Figure~\ref{fig:FermiArc}. It is formed by a pair of planes $\mathcal{S}_1$ and $\mathcal{S}_2$ perpendicular to the $k_z$ axis, together with the side boundaries of the BZ. In the periodic gauge, any BZ slice plane is itself a closed surface (Section~\ref{subsec:Bloch-fns}, Figure~\ref{fig:BZ}), and so has a Chern number in accordance with the Chern theorem, Equation~(\ref{eq:Chern_thm}). The Chern numbers $C_{1,2}$ of the $\mathcal{S}_{1,2}$ slices are defined with respect to a normal vector that points along $k_z$.

When $\mathcal{S}$ encloses the Weyl point at $-\textbf{k}_0$, the Chern number associated with $\mathcal{S}$ for the lower band is $C = 1$.  The side boundaries of the BZ do not contribute to $C$ because the Berry curvature is periodic in the Brillouin zone and the outward normal of one side of the BZ is the inward normal of the opposite side. Hence, $C = C_1 - C_2$. If the bands are topologically trivial for $k_z < -k_0$, then $C_2 = 0$ and $C_1 = C = 1$.

Since the slice $\mathcal{S}_1$ has a non-zero Chern number, it is equivalent to a 2D Chern (QAH) insulator, which in the simplest case has a single edge band crossing $E_\textrm{F}$ (Section~\ref{subsec:QAH}). If the position of $\mathcal{S}_1$ is varied, $C_1$ remains constant as long as $\mathcal{S}$ encloses only the $-\textbf{k}_0$ Weyl point. Hence, $C_1 = 1$ for $-k_0 < k_z < k_0$, and $C_1 = 0$ otherwise. The region between the two Weyl points, therefore, can be viewed as a stack of such Chern insulators, each of which has an edge state that crosses $E_\textrm{F}$.

The detailed dispersion of the edge bands depends on the surface Hamiltonian and boundary conditions. Assuming these are known, one can construct the surface Fermi arcs as follows. Pick a $k_z$, and draw a dot on the surface BZ where the edge band for the slice $\mathcal{S}(k_z)$ crosses $E_\textrm{F}$. Repeat for a series of $k_z$ in the range $-k_0 < k_z < k_0$, and join the dots together to form the arc. The Fermi arc begins and ends on the projection of the Weyl points onto the surface BZ. This is because, firstly, the Weyl points are located at $E_\textrm{F}$ and must therefore lie on the Fermi arc, and secondly, there can be no surface states for $k_z < -k_0$ or $k_z > k_0$ where the bands are topologically trivial. Note that the electronic dispersion on the bottom edge of a Chern insulator is the reverse of that on the top edge (Figure~\ref{fig:QAH}), with the result that the Fermi arcs on opposite surfaces curve in opposite directions (Figure~\ref{fig:FermiArc}).

\subsection{Other characteristics of band crossings}\label{sec:Other}

\begin{figure}
\vspace*{-3cm}
\setlength{\abovecaptionskip}{-60pt plus 0pt minus 0pt}
\centering
\includegraphics[width=0.6\textwidth, angle=90]{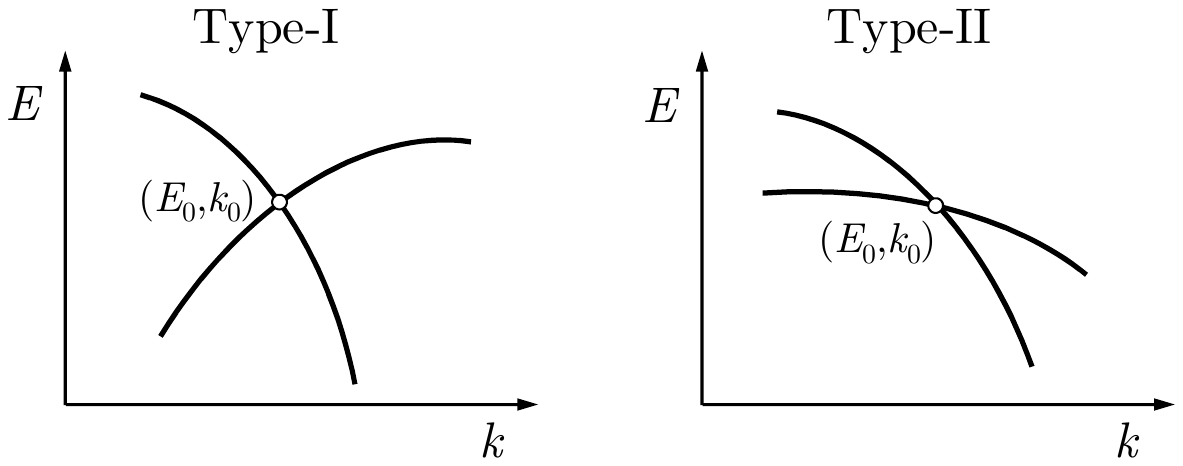}\vspace{0cm}
\caption{Band dispersion along one component of \textbf{k} in the vicinity of Type-I and Type-II nodal points. } \label{fig:WPtypes}
\end{figure}

Before concluding, I want to mention several other differences between the relativistic fermions envisaged in particle physics\footnote{Weyl fermions, incidentally, have not been observed in high-energy physics. Neutrinos were initially believed to be Weyl fermions, but the discovery of neutrino mass put an end to that possibility.} and the emergent relativistic fermions which live in crystalline solids. These differences arise because electrons in crystal lattices are not constrained by the full Poincar\'{e} symmetry of free space.

First, according to Special Relativity, massless particles have a property known as \emph{Lorentz invariance}, which means that they propagate at the speed of light in all directions.  In real band structures, as already mentioned, the Dirac cone is usually tilted and anisotropic (graphene is a notable exception). If the cone is tilted, the group velocities in the directions $+\textbf{q}$ and $-\textbf{q}$ are different (see Figure~\ref{fig:Tilted}). Hence, Lorentz invariance can be broken.

Second, Weyl and Dirac points can be divided into two classes, called type-I and type-II, depending on the degree of tilt, as illustrated in Figure~\ref{fig:WPtypes}. If the tilt is small, the density of states vanishes at the node, and constant-energy surfaces close to the node are ellipsoids. This is a type-I point. If, however, there exists a direction $\textbf{q}$ in which the tilt is sufficiently large that the sign of the gradient of both bands $E^\pm(\textbf{q})$ is the same, either both positive or both negative, then it is defined as a type-II point. Constant-energy surfaces near a type-II point are open, and as a result, the density of states does not vanish at the node. Type-I and type-II topological semimetals have different transport and thermodynamic properties \cite{Soluyanov2015}.

\begin{figure}
\vspace*{-4cm}
\centering
\includegraphics[width=0.75\textwidth, angle=90]{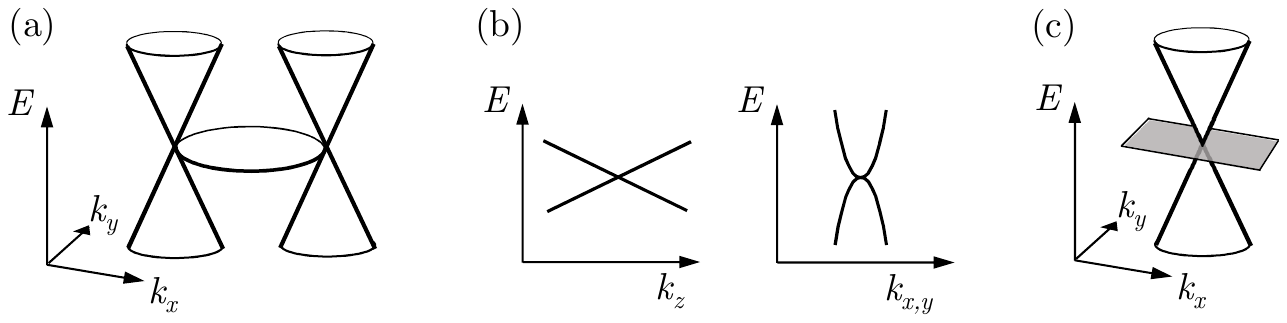}\vspace{0cm}
\setlength{\abovecaptionskip}{-90pt plus 0pt minus 0pt}
\caption{Schematic diagrams of (a) a nodal line, (b) a quadratic Weyl point, with a linear dispersion along $k_z$ and quadratic dispersions along $k_x$ and $k_y$, and (c) a three-fold band crossing.} \label{fig:Nodal}
\end{figure}

Third, under certain circumstances two bands can touch along a line or closed loop in \textbf{k} space, rather than at a point, as sketched in Figure~\ref{fig:Nodal}(a). The curve along which the bands touch is called a nodal line, and if the nodal line energy coincides with $E_\textrm{F}$, the system is termed a \emph{nodal line semimetal}. Both twofold (Weyl) and fourfold (Dirac) nodal lines are possible. There are several different symmetry combinations that can protect a nodal line, and experimental signatures of nodal line semimetals have been predicted \cite{Fang2016,Yang2018}.

Fourth, it is possible to create twofold degenerate band crossings which have a quadratic or cubic dispersion in two directions of \textbf{k} space, and a linear dispersion in the third direction \cite{Fang2012}, Figure~\ref{fig:Nodal}(b). These \emph{multi-Weyl nodes} can occur on lines of symmetry with either 4-fold or 6-fold rotational symmetry in \textbf{k} space.  Closed surfaces surrounding quadratic and cubic Weyl nodes have Chern numbers $C=\pm 2$ and $\pm 3$, respectively.

Finally, in addition to the two- or four-band crossings that define Weyl and Dirac points, three-, six- and eight-band crossings are also possible \cite{Bradlyn2016}, as shown in Figure~\ref{fig:Nodal}(c).  The excitations associated with these points correspond to exotic fermionic quasiparticles with spin representations that differ from one-half and have no counterpart in high-energy physics.

\section{Concluding remarks}\label{sec:Conclusion}

This review covers the basic principles behind the new topological way of thinking about electronic structure in crystalline solids. The central ideas of Berry-ology --- the Berry phase, connection and curvature --- are outlined from their quantum-mechanical origins, and applied to the Bloch states which describe non-interacting electrons in solids. This leads to a topological classification of bulk states in terms of the Chern number and $\mathcal{Z}_2$ topological invariants, and hence to the bulk--boundary correspondence, the wonderful notion that at the boundary between two materials whose electronic states are described by different topological invariants there must inevitably exist conducting surface or edge states.

There are many other important and contemporary topics besides those covered here. These include anomalous charge and thermal transport effects \cite{Armitage2018,XiaoChiangNiu2010,Lv2021}, optically induced transport effects \cite{Lv2021}, a range of magnetoelectric phenomena \cite{Sekine2021,Tokura2019}, the Berry phase theory of electrical polarisation and orbital magnetism \cite{Vanderbilt}, topological superconductivity \cite{SatoAndo2017,Bernevig}, and proposals for topological quantum computation based on Majorana zero modes and other non-Abelian anyons \cite{AguadoKouwenhoven2020,MoessnerMoore,Nayak2008}.

The field of topological materials is a vibrant one, and continues to be fuelled by a wealth of predictions of new quantum states and spin transport phenomena with potential applications in areas such as spintronics, topological electronics and quantum technology. Currently, many of the predictions are as-yet unverified experimentally, and a major effort is underway to identify and synthesise suitable compounds that realise the predicted characteristics. In recent years, this effort has been considerably assisted by the development of publicly available and searchable databases of topological materials \cite{Zhang2019,Vergniory2019,Xu2020,Vergniory2022}.

\section*{Notes on contributor}
\begin{figure}
\vspace*{-3.5cm}
\includegraphics[width=0.5\textwidth, angle=0]{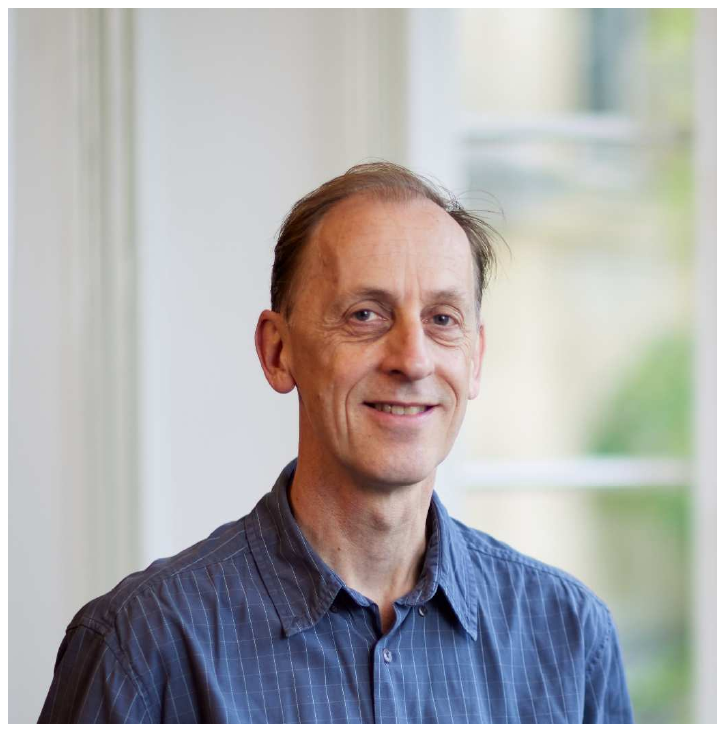}\vspace{0cm}
\vspace*{-2.5cm}
\end{figure}

\textit{Andrew Boothroyd} is a Professor of Physics at Oxford University, and a Tutorial Fellow of Oriel College. Currently he is serving as Associate Head of Department. After completing his undergraduate and graduate degrees at the University of Cambridge, Andrew joined the University of Warwick. In 1992, he moved to Oxford University where he has served as Head of Condensed Matter Physics, Associate Head of Department, and as Vice-Provost of Oriel College. His research exploits neutron and synchrotron X-ray scattering methods to study crystalline, electronic and magnetic order and dynamics in quantum materials.  Recent work has investigated the interplay between magnetic structure and electronic topology in magnetic topological semimetals. In 2011, Andrew received the Institute of Physics’ Brian Pippard Prize, and in 2017 he was recognised as an Outstanding Referee by the American Physical Society.

\section*{ORCID}

\textit{Andrew Boothroyd}: https://orcid.org/0000-0002-3575-7471

\section*{Acknowledgements}

I am grateful to the Laboratory for Neutron Scattering and Imaging at the Paul Scherrer Institute, Switzerland, and the LINXS Institute of Advanced Neutron and X-ray Science, Sweden, for their hospitality and support during extended visits in 2023. I would like to thank Jian-Rui Soh, Fernando de Juan, Maia Vergniory, D. Prabhakaran and Yanfeng Guo for discussions and fruitful collaborations on magnetic topological materials, and Siobhan Tobin and Stephen Blundell for helpful comments on the manuscript.

\section*{Disclosure statement}

No potential conflict of interest was reported by the author.

\section*{Funding}

Support is acknowledged from the Oxford--ShanghaiTech Collaboration Project and from Oriel College, Oxford.

\bibliographystyle{tfnlm}
\bibliography{ElectronicTopology}

\end{document}